\documentclass[12pt]{article}
\usepackage{amscd}
\usepackage{verbatim}
\usepackage{amssymb}
\usepackage{amsmath}
\usepackage{tabularx}
\newcommand{\be}{\begin{equation}}
\newcommand{\ee}{\end{equation}}
\newcommand{\ben}{\begin{eqnarray}}
\newcommand{\een}{\end{eqnarray}}
\newcommand{\ba}{\begin{eqnarray}}
\newcommand{\ea}{\end{eqnarray}}

\newcommand{\bi}{\begin{itemize}}
\newcommand{\ei}{\end{itemize}}

\begin{document}
\begin{center}
\vspace{24pt} { \large \bf Nonlinear perturbations of higher dimensional anti-de Sitter spacetime} \\
\vspace{30pt}
\vspace{30pt}
\vspace{30pt}
{\bf Dhanya S.Menon\footnote{dhanya.menon@students.iiserpune.ac.in}}, {\bf Vardarajan
Suneeta\footnote{suneeta@iiserpune.ac.in}}\\
\vspace{24pt} 
{\em  The Indian Institute of Science Education and Research (IISER),\\
Pune, India - 411008.}
\end{center}
\date{\today}
\bigskip
\begin{center}
{\bf Abstract}
\end{center}
We study nonlinear gravitational perturbations of vacuum Einstein equations, with $\Lambda<0$ in $(n+2)$ dimensions, with $n>2$, generalizing previous studies for $n=2$.  We follow the formalism by Ishibashi, Kodama and Seto to decompose the metric perturbations into tensor, vector and scalar sectors, and simplify the Einstein equations. The tensor perturbations are the new feature of higher dimensions. We render the metric perturbations asymptotically anti-de Sitter by employing a suitable gauge choice for each of the sectors. Finally, we analyze the resonant structure of the perturbed equations at second order for the five dimensional case, by a partial study of single mode tensor-type perturbations at the linear level. For the cases we studied, resonant terms vanish at second order.
\newpage

\section{Introduction}
The stability of the three maximally symmetric solutions to the vacuum Einstein equations
(with cosmological constant) has been studied extensively. Minkowski and de Sitter
spacetime have been found to be nonlinearly stable under small perturbations \cite{christ}, \cite{friedrich}. Although Anti de Sitter (AdS) spacetime is
stable under linearized perturbations, its nonlinear instability was only discovered recently in
a seminal work \cite{bizon} by Bizon and Rostworowski. They did a numerical study, which
involved evolution of a massless scalar field in four
dimensions with spherical symmetry and with cosmological constant $\Lambda<0$. The end point of the evolution was a black
hole, indicating the nonlinear instability of the system. These results
hold true for higher dimensions as well \cite{jalmuzna}.

The AdS instability was also observed in the evolution of
complex scalar fields \cite{liebling1}. The evolution of
a massless scalar field in Gauss Bonnet gravity was numerically studied in
\cite{kunstatter}, \cite{kunstatter1}. Further, renormalization group methods \cite{craps1},
\cite{craps2} and the two-time framework (TTF) \cite{balas},\cite{green},
\cite{maliborski6} were used to study the instability of AdS.
An interacting scalar field in AdS was
investigated in TTF in \cite{basu}.

Certain systems like a
massless scalar field enclosed in a cavity in Minkowski
\cite{maliborski} and massive fields in AdS \cite{pani}, were thought to
exhibit an  AdS-like
instability, although the linear spectra were non-resonant. This led to the question of whether resonant spectra are required for a
turbulent instability. It was later seen that there is a minimum amplitude
required to
trigger instability in such cases \cite{maliborski3}. This minimum
amplitude was too small to
be observed initially in numerical studies. The reasons for these observations were discussed in
\cite{ads}.

Non collapsing solutions were studied for asymptotically AdS spacetimes in \cite{maliborski3},
\cite{maliborski1}, \cite{buchel},
\cite{maliborski5}, \cite{arias}, \cite{kim} and \cite{fodor}.

Finally, the proof of AdS instability for the Einstein-massless
Vlasov system in spherical symmetry was given by Moschidis \cite{moschidis}.

Non spherically symmetric collapse was also studied in \cite{bantilan}
with massless scalar field in five dimensions. It was seen that the
configuration collapsed faster than the spherically symmetric case.
Rotational dynamics of $AdS_5$ was studied in \cite{choptuik} in presence
of a complex doublet scalar field.

The study of pure gravitational perturbations naturally involves breaking of
spherical symmetry. Numerical evolution for pure gravity
was done in $AdS_5$ using the cohomogeneity-two biaxial Bianchi IX ansatz
\cite{cohomo}. A more general breakdown of spherical symmetry was done in
$AdS_4$
\cite{horowitz}, where time periodic solutions called geons were
constructed. Geons were also constructed in \cite{horowitz1},
\cite{rost}, \cite{rost2}, \cite{diasantos}, \cite{martinon2}, \cite{rost3}
and \cite{fodor1}. Nonlinear perturbation theory was employed to study pure
gravitational perturbations in a cavity in Minkowski, in general dimensions
\cite{mincav}. Very recently, the resonant system in five dimensions with cohomogenity-two biaxial Bianchi IX ansatz was discussed in \cite{adsbi5}.

In this work, we generalize the study of gravitational
perturbations of AdS using nonlinear perturbation theory in four
dimensions \cite{rost} to $(n+2)$
dimensions $(n>2)$. In section II, we give an overview of the methods used
to study
pure gravitational perturbations of $ AdS_{n+2}$. We use the
Kodama-Ishibashi formalism \cite{seto} to study nonlinear perturbations, by
extending it beyond the linear level. Section III holds a brief discussion
on how the metric perturbations and hence the source terms fall off,
upon imposing asymptotic AdS boundary conditions. Section IV contains
simplification
of higher order equations. In section V-VII we systematically construct
metric perturbations and render them asymptotically AdS through suitable
gauge choices. Section VIII
contains calculation of $l_s=0,1$ scalar modes as well as $l_v=1$ vector
modes. These modes, which are gauge at linear level, are in fact physical
perturbations at higher orders. In Section IX, we study an example of single mode tensor perturbations as initial data and investigate if secular terms arise at second order in dimension five ($n=3$). For the cases considered, the secular terms vanish. Finally, section X contains the summary
and discussion of this paper.

\section{Methodology}
The vacuum Einstein equation with negative cosmological constant (i.e.
$\Lambda<0 )$ in $(n+2)$ spacetime dimensions is given by
\begin{align}
R_{\mu\nu}+\frac{(n+1)}{L^2} g_{\mu\nu}=0
\label{eq1}
\end{align}
where $L^2=-\frac{n(n+1)}{2\Lambda}$. We are interested in the solutions of
the above equation when the metric $g_{\mu\nu}$ is expanded around the AdS
metric, which
we will refer to as the background metric, given by
\begin{align}
ds^2=-f(r)
dt^2+f(r)^{-1}dr^2+r^2d\Omega^2_{n}~~,f(r)=1+\frac{r^2}{L^2}\label{eq2}
\end{align}
where $d\Omega_n^2=\gamma_{ij}(w)dw^idw^j$
is the  metric for $n-$sphere.
We are interested in
generic perturbations about the background metric to higher orders in perturbation theory. In
the equations that follow, we employ the notation used by \cite{rost}.
Hence the `bar' quantities refer to the background AdS geometry. Next we
expand the solutions of (\ref{eq1}) as $g_{\mu\nu}=\bar{g}_{\mu\nu}+\delta
g_{\mu\nu}$, where
\begin{align}
\delta g_{\mu\nu}=\sum_{1\leq i}{^{(i)}} h_{\mu\nu}\epsilon^i
\label{eq3}
\end{align}
Then the inverse metric can be represented as
\begin{align}
g^{\alpha\beta}&=(\bar{g}^{-1}-\bar{g}^{-1}\delta g
\bar{g}^{-1}+\bar{g}^{-1}\delta g \bar{g}^{-1}\delta g
\bar{g}^{-1}-...)^{\alpha\beta}\nonumber\\&=\bar{g}^{\alpha\beta}+\delta
g^{\alpha\beta}.
\label{eq4}
\end{align}
The Christoffel symbol as:
\begin{align}
\Gamma^{\alpha}_{\mu\nu}&=\bar{\Gamma}^{\alpha}_{\mu\nu}+\frac{1}{2}(\bar{g}^{-1}-\bar{g}^{-1}\delta
g \bar{g}^{-1}+...)^{\alpha\beta}( \bar{\nabla}_{\mu}\delta
g_{\beta\nu}+\bar{\nabla}_{\nu}\delta
g_{\beta\mu}-\bar{\nabla}_{\beta}\delta g_{\mu\nu}
)\nonumber\\&=\bar{\Gamma}^{\alpha}_{\mu\nu}+\delta
\Gamma^{\alpha}_{\mu\nu},
\label{eq5}
\end{align}
And the Ricci tensor as:
\begin{align}
R_{\mu\nu}&=\bar{R}_{\mu\nu}+\bar{\nabla}_{\alpha}\delta
\Gamma^{\alpha}_{\mu\nu}-\bar{\nabla}_{\nu}\delta
\Gamma_{\alpha\mu}^{\alpha}+\delta\Gamma^{\alpha}_{\alpha\lambda}\delta
\Gamma^{\lambda}_{\mu\nu}-\delta \Gamma^{\lambda}_{\mu\alpha}\delta
\Gamma^{\alpha}_{\lambda\nu}\nonumber\\&=\bar{R}_{\mu\nu}+\delta R_{\mu\nu}.
\label{eq6}
\end{align}
From here onwards, similar to $\delta g_{\mu\nu}$, the various perturbed quantities will be written as an expansion in powers of parameter $\epsilon^i$. In such a case, the superscript $(i)$ on the left hand side of a quantity denotes the order of perturbation theory we are in.
Hence, the perturbed Einstein's equation is given by
\begin{align}
{^{(i)}}\!R_{\mu\nu}+\frac{(n+1)}{L^2}{^{(i)}}\!h_{\mu\nu}=0
\label{eq7}
\end{align}
Before writing down our working equations, we define the following two
quantities:
The Lorentzian Lichnerowicz operator $\triangle_L$, which is given as
\begin{align}
2\triangle_L
{^{(i)}}h_{\mu\nu}=&-\bar{\nabla}^{\alpha}\bar{\nabla}_{\alpha}{^{(i)}}h_{\mu\nu}-\bar{\nabla}_{\mu}\bar{\nabla}_{\nu}{^{(i)}}h+\bar{\nabla}_{\mu}\bar{\nabla}_{\alpha}{^{(i)}}h^{\alpha}_{\nu}+\bar{\nabla}_{\nu}\bar{\nabla}_{\alpha}{^{(i)}}h^{\alpha}_{\mu}\nonumber\\&+\bar{R}_{\mu\alpha}{^{(i)}}h^{\alpha}_{\nu}+\bar{R}_{\nu\alpha}
{^{(i)}}h^{\alpha}_{\mu}-2\bar{R}_{\mu\alpha\nu\lambda}{^{(i)}}h^{\alpha\lambda}
\label{eq8}
\end{align}
and  ${^{(i)}}\!A_{\mu\nu}$ defined as
\begin{align}
{^{(i)}}\!A_{\mu\nu}=&[\epsilon^i]\Big\{-\bar{\nabla}_{\alpha}\big[
(-\bar{g}^{-1}\delta
g\bar{g}^{-1}+...)^{\alpha\lambda}(\bar{\nabla}_{\mu}\delta
g_{\lambda\nu}+\bar{\nabla}_{\nu}\delta
g_{\lambda\nu}-\bar{\nabla}_{\lambda}\delta g_{\mu\nu})\big]
\nonumber\\&+\bar{\nabla}_{\nu}\big[ (-\bar{g}^{-1}\delta
g\bar{g}^{-1}+...)^{\alpha\lambda}(\bar{\nabla}_{\mu}\delta
g_{\lambda\alpha}+\bar{\nabla}_{\alpha}\delta
g_{\lambda\mu}-\bar{\nabla}_{\lambda}\delta g_{\mu\alpha})\big]\nonumber\\&-
2\delta\Gamma^{\alpha}_{\alpha\lambda}\delta\Gamma^{\lambda}_{\mu\nu}+2\delta\Gamma^{\lambda}_{\mu\alpha}\delta\Gamma^{\alpha}_{\lambda\nu}
\Big\}
\label{eq9}
\end{align}
where $[\epsilon^i] \; f $ denotes the coefficient of $\epsilon^i$  in the
expansion of the power series $\sum_i \epsilon^i f_i$.
Next, we derive the working equation.
Note that $2{^{(i)}}\!R_{\mu\nu}=2\triangle_L{^{(i)}}\!h_{\mu\nu}-{^{(i)}}\!A_{\mu\nu}$. Hence, the perturbed Einstein equation (\ref{eq7}) takes the form
\begin{align}
2\triangle_L{^{(i)}}\!h_{\mu\nu}+\frac{2(n+1)}{L^2}{^{(i)}}\!h_{\mu\nu}={^{(i)}}\!A_{\mu\nu}
\label{eqa3}
\end{align}
 We will now take the trace of this equation and divide it by 2. Since the background AdS metric also obeys (\ref{eq1}), we also make the replacement $\frac{(n+1)}{L^2}\bar{g}^{\alpha\beta}=-\bar{R}^{\alpha\beta}
$, so as to obtain
\begin{align}
\bar{g}^{\alpha\beta}\triangle_L{^{(i)}}\!h_{\alpha\beta}-\bar{R}^{\alpha\beta}{^{(i)}}\!h_{\alpha\beta}   =\frac{1}{2}\bar{g}^{\alpha\beta}{^{(i)}}\!A_{\alpha\beta}
\label{eqa4}
\end{align}
Subtracting $\bar{g}_{\mu\nu}$(\ref{eqa4}) from (\ref{eqa3}) gives us
\begin{align}
2\triangle_L{^{(i)}}\!h_{\mu\nu}+\frac{2(n+1)}{L^2}{^{(i)}}\!h_{\mu\nu}-\bar{g}_{\mu\nu}(\bar{g}^{\alpha\beta}\triangle_L{^{(i)}}\!h_{\alpha\beta}&-{^{(i)}}\!h_{\alpha\beta}\bar{R}^{\alpha\beta})\nonumber\\&={^{(i)}}\!A_{\mu\nu}-\frac{1}{2}\bar{g}_{\mu\nu}\bar{g}^{\alpha\beta}{^{(i)}}\!A_{\alpha\beta}
\label{eqa5}
\end{align}
which we rewrite as
\begin{align}
{^{(i)}}\!G_{\mu\nu}=2\tilde{\triangle}_L{^{(i)}}\!h_{\mu\nu}-{^{(i)}}\!S_{\mu\nu}=0\label{eq10}
\end{align}
where $\tilde{\triangle}_L{^{(i)}}\!h_{\mu\nu}$ is given by
\begin{align}
2\tilde{\triangle}_L{^{(i)}}\!h_{\mu\nu}=2\triangle_L{^{(i)}}\!h_{\mu\nu}+\frac{2(n+1)}{L^2}{^{(i)}}\!h_{\mu\nu}-\bar{g}_{\mu\nu}(\bar{g}^{\alpha\beta}\triangle_L{^{(i)}}\!h_{\alpha\beta}-{^{(i)}}\!h_{\alpha\beta}\bar{R}^{\alpha\beta})
\label{eq11}
\end{align}
and ${^{(i)}}\!S_{\mu\nu}$, which is given in terms of ${^{(i)}}\!A_{\mu\nu}$ is
defined as
\begin{align}
{^{(i)}}\!S_{\mu\nu}={^{(i)}}\!A_{\mu\nu}-\frac{1}{2}\bar{g}_{\mu\nu}\bar{g}^{\alpha\beta}{^{(i)}}\!A_{\alpha\beta}
\label{eq12}
\end{align}
The background metric $\bar{g}_{\mu\nu}$ is spherically symmetric and is of
the form
\begin{align}
ds^2=\bar{g}_{\mu\nu}dz^{\mu}dz^{\nu}=g_{ab}(y)dy^ady^b+r^2(y)d\Omega_n^2
\label{eq13}
\end{align}
 One can use the gauge invariant formalism given by Ishibashi, Kodama and Seto
\cite{seto} to study the perturbations around such a background metric and we will be extending the same to  higher orders in perturbations theory as well. Let the
covariant derivative associated with $ds^2$, $g_{ab}dy^ady^b$ and
$d\Omega_n^2$ be $\bar{\nabla}_M$, $\bar{D}_a$ and  $\bar{D}_i$
respectively. The metric perturbations
${^{(i)}}h_{\mu\nu}$ are decomposed according to their behaviour on the
$n-$sphere i.e.
into the  scalar type, $\mathbb{S}$, the vector type, $\mathbb{V}_i$  and
the tensor type, $\mathbb{T}_{ij}$. In the following sections,
$\hat{\vartriangle}=\hat{D}^i\hat{D}_i$ where raising (and lowering) of
the sphere indices is done with $\gamma_{ij}$.
The scalar harmonics $\mathbb{S}$ satisfy
\begin{align}
(\hat{\vartriangle}+k^2_s)\mathbb{S}_{\textbf{k}_s}=0
\label{eq15}
\end{align}
where $k_s^2=l_s(l_s+n-1)$ and $l_s=0,1,...$ Also, $\textbf{k}_s$ is the  multi-index of the form $\{l_s,l_s^{(1)}...l_s^{(n-1)}=m_s  \}$, where $l_s, l_s^{(1)}...$ denote the various quantum numbers, such that $l_s\geq l_s^{(1)}\geq l_s^{(2)}\ldots \geq l_s^{(n-2)} \geq |m_s|$. From $\mathbb{S}$, one can construct $\mathbb{S}_i$ and $\mathbb{S}_{ij}$
\begin{align}
\mathbb{S}_i=-\frac{1}{k_s}\bar{D}_i\mathbb{S};~~
\mathbb{S}_{ij}=\frac{1}{k^2_s}\bar{D}_i\bar{D}_j\mathbb{S}+\frac{1}{n}\gamma_{ij}\mathbb{S}
\label{eq16}
\end{align}
which satisfy
\begin{align}
\bar{D}^i\mathbb{S}_i=k_s\mathbb{S};~~
{\mathbb{S}_i^i=0;~~\bar{D}_j\mathbb{S}^j_i=\frac{(n-1)(k^2_s-n)}{nk_s}\mathbb{S}_i}.
\label{eq19}
\end{align}
Vector harmonics $\mathbb{V}_i$ are defined as
\begin{align}
(\hat{\vartriangle}+k^2_v)\mathbb{V}_{\textbf{k}_vi}=0
\label{eq20}
\end{align}
where $k^2_v=l_v(l_v+n-1)-1$ and $l_v=1,2,...$, such that
\begin{align}
{\bar{D}_i\mathbb{V}^i=0}.
\label{eq21}
\end{align}
 Here, $\textbf{k}_v$ is the multi-index associated with vector harmonics. From $\mathbb{V}_i$, one can construct tensors $\mathbb{V}_{ij}$
\begin{align}
\mathbb{V}_{ij}=-\frac{1}{2k_v}(\bar{D}_i\mathbb{V}_j+\bar{D}_j\mathbb{V}_i)
\label{eq22}
\end{align}
which satisfy
\begin{align}
\mathbb{V}_i^i=0;~~
\bar{D}_j\mathbb{V}_i^j=\frac{(k^2_v-(n-1))}{2k_v}\mathbb{V}_i
\label{eq23}
\end{align}
Tensor type harmonics, $\mathbb{T}_{ij}$ are defined as
\begin{align}
(\hat{\vartriangle}+k^2)\mathbb{T}_{\textbf{k}ij}=0
\label{eq24}
\end{align}
where $k^2=l(l+n-1)-2$ and $l=2,3,...$. $\textbf{k}$ denotes the multi-index associated with tensor harmonics. They satisfy
\begin{align}
\mathbb{T}_i^i=0 ;~~\bar{D}_j\mathbb{T}_i^j=0
\label{eq25}
\end{align}
The metric perturbations can be now be expanded as
\begin{eqnarray}
{^{(i)}}h_{ab}=\sum_{\bf{k}_s}{^{(i)}}\!f_{ab\bf{k}_s}\mathbb{S}_{\bf{k}_s}
; ~~
{^{(i)}}h_{ai}=r\bigg(\sum_{\bf{k}_s}{^{(i)}}\!f_{a\bf{k}_s}^{(s)}\mathbb{S}_{\textbf{k}_si}+\sum_{\bf{k}_v}{^{(i)}}\!f_{a\bf{k}_v}^{(v)}\mathbb{V}_{\textbf{k}_vi}\bigg)\nonumber
\end{eqnarray}
\begin{align}
{^{(i)}}h_{ij}=&r^2\bigg(\sum_{\textbf{k}}{^{(i)}}\!H_{T\textbf{k}}\mathbb{T}_{\textbf{k}ij}+2\sum_{\textbf{k}_v}{^{(i)}}H^{(v)}_{T\textbf{k}_v}\mathbb{V}_{\textbf{k}_vij}\nonumber\\&+2\sum_{\textbf{k}_s}(
{^{(i)}}\!H^{(s)}_{T\textbf{k}_s}\mathbb{S}_{\textbf{k}_sij}+{^{(i)}}H_{L\textbf{k}_s}\gamma_{ij}\mathbb{S}_{\textbf{k}_s})\bigg)
\label{eq26}
\end{align}
The metric components are also gauge dependent. Under an infinitesimal
gauge transformation
$\bar{\delta}z^{\alpha}=\sum_{i}{^{(i)}}\zeta^{\alpha}\epsilon^i$, metric
perturbation ${^{(i)}}h_{\mu\nu}$ transforms as
\begin{align}
{^{(i)}}h_{\mu\nu}\rightarrow
{^{(i)}}h_{\mu\nu}-\bar{\nabla}_{\mu}{^{(i)}}\zeta_{\nu}-\bar{\nabla}_{\nu}{^{(i)}}\zeta_{\mu}
\label{eq27}
\end{align}
i.e
\begin{align}
{^{(i)}}h_{ab}\rightarrow
&{^{(i)}}h_{ab}-\bar{D}_a{^{(i)}}\zeta_b-\bar{D}_b{^{(i)}}\zeta_a\nonumber\\
{^{(i)}}h_{ai}\rightarrow
&{^{(i)}}h_{ai}-\bar{D}_i{^{(i)}}\zeta_a-r^2\bar{D}_a\left(
\frac{{^{(i)}}\zeta_i}{r^2}\right)\nonumber\\
{^{(i)}}h_{ij}\rightarrow &
{^{(i)}}h_{ij}-\bar{D}_i{^{(i)}}\zeta_j-\bar{D}_j{^{(i)}}\zeta_i\nonumber\\&-2r\bar{D}^ar\:{^{(i)}}\zeta_a\gamma_{ij}
\label{eq28}
\end{align}
Let ${^{(i)}}\zeta_a={^{(i)}}T_a\mathbb{S}$ and
${^{(i)}}\zeta_i=r\:{^{(i)}}\!M\mathbb{S}_i+r\:{^{(i)}}\!M^{(v)}\:\mathbb{V}_i$.
Thus the gauge transformations for ${^{(i)}}f_{ab}$, ${^{(i)}}f_a^{(s)}$,
${^{(i)}}f_a^{(v)}$, ${^{(i)}}H_T^{(s)}$, ${^{(i)}}H_T^{(v)}$,
${^{(i)}}H_L$ and ${^{(i)}}H_T$ are
\begin{align}
{^{(i)}}f_{ab}&\rightarrow
{^{(i)}}f_{ab}-\bar{D}_a{^{(i)}}T_b-\bar{D}_b{^{(i)}}T_a\\
{^{(i)}}f^{(s)}_a &\rightarrow {^{(i)}}f^{(s)}_a-r\bar{D}_a\left(
\frac{{^{(i)}}M}{r}\right)+\frac{k_s}{r}{^{(i)}}T_a\\
{^{(i)}}H_L &\rightarrow
{^{(i)}}H_L-\frac{k_s}{nr}{^{(i)}}M-\frac{\bar{D}^a r}{r}{^{(i)}}T_a\\
{^{(i)}}H_T^{(s)}&\rightarrow {^{(i)}}H_T^{(s)}+\frac{k_s}{r}{^{(i)}}M\\
{^{(i)}}f_a^{(v)}&\rightarrow{^{(i)}}
f_a^{(v)}-r\bar{D}_a\left(\frac{{^{(i)}}M^{(v)}}{r}\right)\\
{^{(i)}}H_T^{(v)}&\rightarrow
{^{(i)}}H_T^{(v)}+\frac{k_v}{r}{^{(i)}}M^{(v)}\\
{^{(i)}}H_T &\rightarrow {^{(i)}}H_T
\label{eq29}
\end{align}
For all cases except $l_s=0,1$ and $l_v=1$ modes, one can define the
following gauge invariant variables.
\begin{align}
{^{(i)}}\!Z_a={^{(i)}}\!f_a^{(v)}+\frac{r}{k_v}\bar{D}_a{^{(i)}}\!H_T^{(v)}
\label{eq30}
\end{align}
\begin{align}
{^{(i)}}F_{ab}={^{(i)}}\!f_{ab}+\frac{1}{2}\bar{D}_{(a}{^{(i)}}\!X_{b)};~~{^{(i)}}\!F={^{(i)}}H_L+\frac{{^{(i)}}\!H_T^{(s)}}{n}+\frac{1}{r}\bar{D}^ar{^{(i)}}\!X_a
\label{eq31}
\end{align}
where
\begin{align}
{^{(i)}}X_a=\frac{r}{k_s}\left({^{(i)}}\!f_a^{(s)}+\frac{r}{k_s}\bar{D}_a{^{(i)}}\!H_T^{(s)}\right)
\label{eq32}
\end{align}
It is possible to write the $\tilde{\triangle}_L$ operator in (\ref{eq10})
solely in terms of these gauge invariant variables \cite{seto}. The strategy is to
solve for these variables and add suitable gauge transformations to the
metric perturbations to render them asymptotically AdS (aAdS).

\section{Asymptotic nature of source terms}
The metric perturbations ${^{(i)}}\!h_{\mu\nu}$ are dependent on the sources ${^{(i)}}\!S_{\mu\nu}$ as  well. The sources in turn are constructed from the metric perturbations of the previous orders. In this section, we will deduce the leading order behaviour of ${^{(i)}}\!S_{\mu\nu}$ as $r\rightarrow\infty$. This is essential so that we can fix the gauge appropriately so as to render the metric perturbations asymptotically AdS.

The metric perturbations $\delta g_{\mu\nu}$ satisfing asymptotically AdS conditions will have
the following leading order behaviour as $r\rightarrow\infty$ \cite{Hennaux}, \cite{hennauhigher}, \cite{hennaucharge}:
\begin{align}
\delta g_{rr}\sim\frac{1}{r^{n+3}}~~;\delta
g_{r\gamma}\sim\frac{1}{r^{n+2}}~~;\delta
g_{\gamma\sigma}\sim\frac{1}{r^{n-1}}
\label{eq33}
\end{align}
where $\sigma,\gamma\neq r$.
Given this, we deduce the fall off of $\delta g^{\mu\nu}$ as $r\rightarrow\infty$.
Equation (\ref{eq33}) tells us that as $r\rightarrow\infty$, each of the ${^{(i)}}\!h^{\mu\nu}=\bar{g}^{\mu\alpha}\bar{g}^{\nu\beta}{^{(i)}}\!h_{\alpha\beta}$ should  fall off atleast like
\begin{align}
{^{(i)}}\!h^{rr}\sim\frac{1}{r^{n-1}}~~;{^{(i)}}\!h^{r\gamma}\sim\frac{1}{r^{n+2}}~~;{^{(i)}}\!h^{\gamma\sigma}\sim\frac{1}{r^{n+3}}
\label{eqb1}
\end{align}
Similarly, a term like ${^{(i)}}\!h_{\mu}^{\nu}$ falls off at least like $\frac{1}{r^n}$.
Now, in general, the $i-$th order component of $\delta g^{\mu\nu}$, say ${^{(i)}}\!f^{\mu\nu}$ will be of the form
\begin{align}
{^{(i)}}\!f^{\mu\nu}=-{^{(i)}}\!h^{\mu\nu}-\sum_{x=1}^{(i-1)}{^{(x)}}\!h^{\mu\lambda_1}{^{(i-x)}}\!f^{\nu}_{\lambda_1}
\label{eqb2}
\end{align}
Our aim is to prove, that at any order, the fall off of ${^{(i)}}\!f^{\mu\nu}$ is the same as the fall off of ${^{(i)}}\!h^{\mu\nu}$.
For $i=1$, ${^{(1)}}\!f^{\mu\nu}=-{^{(1)}}\!h^{\mu\nu}$ (from here on, we omit the label $(1)$ on linear perturbations),
and hence (\ref{eqb1}) holds true. For $i=2$,
\begin{align}
{^{(2)}}\!f^{\mu\nu}=-{^{(2)}}\!h^{\mu\nu}+h^{\mu\lambda_1}h_{\lambda_1}^{\nu}
\label{eqb3}
\end{align}
From (\ref{eqb3}), one can easily deduce that the leading order term is contributed only by ${^{(i)}}\!h_{\mu\nu}$, since the products of the metric perturbations tend to fall off at a faster rate.

Thus, at any order $i$, the leading order behaviour of ${^{(i)}}\!f^{\mu\nu}$ is the same as ${^{(i)}}\!h^{\mu\nu}$, because the rest of the terms in (\ref{eqb2}) tend to fall off faster than ${^{(i)}}\!h^{\mu\nu}$. For eg., in case of ${^{(i)}}\!f^{rr}$, the terms in $\sum_{x=1}^{(i-1)}{^{(x)}}\!h^{r\lambda_1}\:{^{(i-x)}}\!f^r_{\lambda_1}$ are at least of the order $\frac{1}{r^{2n}}$.

Similarly, in case of ${^{(i)}}\!f^{rt}$, while ${^{(i)}}\!h^{rt}$ falls off like $\frac{1}{r^{n+2}}$, the rest of the terms in $\sum_{x=1}^{i-1}{^{(x)}}\!h^{r\lambda_1}{^{(i-1)}}\!f^{t}_{\lambda_1}$ fall off atleast like $\frac{1}{r^{2n+3}}$. On the same lines, one can deduce that the leading order fall off of ${^{(i)}}\!f^{tt}$ is same as that of ${^{(i)}}\!h^{tt}$, which is $\frac{1}{r^{n+3}}$, because the rest of the terms fall off like $\frac{1}{r^{2n+4}}$ or at a faster rate.
 Hence, the leading order behaviour of   $\delta g^{\mu\nu}$ should be:
\begin{align}
\delta g^{rr}\sim \frac{1}{r^{n-1}} ~~;\delta g^{r\gamma}\sim\frac{1}{r^{n+2}}~~;\delta g^{\gamma\sigma}\sim\frac{1}{r^{n+3}}
\label{eq33.1}
\end{align}
Now, we turn our attention to the behaviour of the sources ${^{(i)}}\!S_{\mu\nu}$ as $r\rightarrow\infty$.

To do this, we first simplify (\ref{eq9}) and then use (\ref{eq33}) and (\ref{eq33.1}) in the resultant expansion. Since the calculations are too
tedious, we present few of the terms, which indicate the leading order
behaviour in ${^{(i)}}\!A_{\mu\nu}$. As a prerequisite, we also give the
leading order fall off of the various Christoffel symbols, defined on the
background AdS metric
\begin{align}
&\bar{\Gamma}_{tr}^t=\frac{f'}{2f}\sim\frac{1}{r}\hspace{2 cm} \bar{\Gamma}_{rr}^r=-\frac{f'}{2f}\sim\frac{1}{r}\nonumber\\&
\bar{\Gamma}_{kr}^k=\frac{n}{r}\sim\frac{1}{r}
\label{eqb5}
\end{align}

Consider a term of the following form in the expansion of ${^{(i)}}\!A_{\mu\nu}$
\begin{align*}
-[\epsilon^i]\bar{\Gamma}^{\alpha}_{\alpha\lambda_1}(-\bar{g}^{-1}\delta
g\bar{g}^{-1}+...)^{\lambda_1\lambda}(-\bar{\nabla}_{\lambda}\delta
g_{\mu\nu}+\bar{\nabla}_{\mu}\delta g_{\nu\lambda} +\bar{\nabla}_{\nu}\delta g_{\mu\lambda})
\end{align*}
One of the components of this term in ${^{(i)}}\!A_{tt}$ is
\begin{align}
&-[\epsilon^i] \bar{\Gamma}_{tr}^t[(-\bar{g}^{-1}\delta
g\bar{g}^{-1}+...)^{rr}(-\bar{\nabla}_r\delta g_{tt} +...  )
]\nonumber\\&\sim \frac{1}{r}\times \frac{1}{r^{n-1}}\times
\frac{1}{r^n}\nonumber\\&
\sim \frac{1}{r^{2n}}
\end{align}
Here, we have used the fact that $(-\bar{g}^{-1}\delta
g\bar{g}^{-1}+...)^{\mu\nu}$ will have the same leading order behaviour as that of $\delta g^{\mu\nu}$.
Similarly, for ${^{(i)}}\!A_{ij}$, one obtains,
\begin{align}
&-[\epsilon^i]\bar{\Gamma}_{kr}^k(-\bar{g}^{-1}\delta g\bar{g}^{-1}+...)^{rr}(-\bar{\nabla}_r\delta
g_{ij}+...)\nonumber\\&
\sim \frac{1}{r}\times\frac{1}{r^{n-1}}\times\frac{1}{r^n}\nonumber\\&
\sim \frac{1}{r^{2n}}
\end{align}
Next, consider another term in ${^{(i)}}\!A_{\mu\nu}$ of the form
\begin{align*}
-[\epsilon^i]\partial_{\alpha}[(-\bar{g}^{-1}\delta g
\bar{g}^{-1}+...)^{\alpha\lambda}(-\bar{\nabla}_{\lambda}\delta
g_{\mu\nu}+\bar{\nabla}_{\mu}\delta g_{\lambda\nu}+\bar{\nabla}_{\nu}\delta
g_{\lambda\mu})]
\end{align*}

In ${^{(i)}}\!A_{ir}$, the above term contains the following part
\begin{align}
&-[\epsilon^i]\partial_t[(-\bar{g}^{-1}\delta g\bar{g}^{-1}+...)^{tt}(\partial_r\delta
g_{it})  ]\nonumber\\&
\sim \frac{1}{r^{n+3}}\times \frac{1}{r^n}\nonumber\\&
\sim \frac{1}{r^{2n+3}}
\end{align}
Similarly, in ${^{(i)}}\!A_{rt}$, following component is present
\begin{align}
&-[\epsilon^i]\partial_{t}[(-\bar{g}^{-1}\delta
g\bar{g}^{-1}+...)^{tt}\bar{\nabla}_r\delta g_{tt} ]\nonumber\\&
\sim \frac{1}{r^{n+3}}\times \frac{1}{r^n}\nonumber\\&
\sim \frac{1}{r^{2n+3}}
\end{align}

Lastly, one of the terms in  ${^{(i)}}\!A_{rr}$ is
\begin{align*}
[\epsilon^i]\bar{\Gamma}_{\alpha r}^{\lambda_1}(-\bar{g}^{-1}\delta
g\bar{g}^{-1})^{\alpha\lambda}[-\bar{\nabla}_{\lambda}\delta
g_{r\lambda_1}+\bar{\nabla}_{r}\delta
g_{\lambda\lambda_1}+\bar{\nabla}_{\lambda_1}\delta g_{r\lambda}]
\end{align*}

From this, the component which contributes to the leading order term is
\begin{align}
&[\epsilon^i]\bar{\Gamma}_{rr}^r(-\bar{g}^{-1}\delta g\bar{g}^{-1}+...)^{rr}\bar{\nabla}_r\delta
g_{rr}\nonumber\\&
\sim\frac{1}{r}\times\frac{1}{r^{n-1}}\times\frac{1}{r^{n+4}}\nonumber\\&
\sim\frac{1}{r^{2n+4}}
\end{align}
Finally, we combine all these results and get the following behaviour for ${^{(i)}}\!S_{\mu\nu}$ as $r\rightarrow\infty$

\begin{align}
{^{(i)}}\!S_{rr}\sim\frac{1}{r^{2n+4}}~~;{^{(i)}}\!S_{r\gamma}\sim\frac{1}{r^{2n+3}}~~;{^{(i)}}\!S_{\gamma\sigma}\sim\frac{1}{r^{2n}}
\label{eq34}
\end{align}

\section{Linear and higher order equations}
The  spherical symmetry of the background makes it possible to write the various background quantities in terms of the connections $\bar{D}_a$ and $\bar{D}_i$, which are defined on $g_{ab}dy^ady^b=-fdt^2+f^{-1}dr^2$ and $d\Omega^2$ respectively \cite{seto}. They are given by (the quantities which are solely defined on $g_{ab}dy^ady^b$ have a superscript "m" on their left hand side whereas a "hat" on a quantity means it is solely defined on the $n-$sphere.):\\
\textbf{Christoffel symbols}
\begin{align}
&\bar{\Gamma}_{bc}^a={^{m}}\!\Gamma_{bc}^a,~~\bar{\Gamma}_{ij}^a=-r\bar{D}^ar\gamma_{ij}\nonumber\\&
\bar{\Gamma}_{aj}^i=\frac{\bar{D}_ar}{r}\delta_j^i,~~\bar{\Gamma}_{jk}^i=\hat{\Gamma}_{jk}^i
\label{lh13}
\end{align}
\textbf{Curvature tensors}
\begin{align}
&\bar{R}^a_{\:bcd}={^{m}}\!R^a_{\:bcd}\nonumber\\&
\bar{R}^i_{\:ajb}=-\frac{\bar{D}_a\bar{D}_br}{r}\delta_j^i\nonumber\\&
\bar{R}^i_{\:jkl}=[1-(\bar{D}r)^2](\delta^i_k\gamma_{jl}-\delta_l^i\gamma_{jk})
\label{lh14}
\end{align}
\textbf{Ricci tensors}
\begin{align}
&\bar{R}_{ab}={^{m}}\!R_{ab}-\frac{n}{r}\bar{D}_a\bar{D}_br\nonumber\\&
\bar{R}_{ai}=0\nonumber\\&
\bar{R}^i_j=\left( -\frac{\bar{D}^a\bar{D}_ar}{r}+(n-1)\frac{(1-(\bar{D}r)^2)}{r^2} \right)\delta_j^i\nonumber\\&
\bar{R}={^{m}}\!R-2n\frac{\bar{D}^a\bar{D}_ar}{r}+n(n-1)\frac{(1-(\bar{D}r)^2)}{r^2}
\label{lh15}
\end{align}
The motive is to write equation (\ref{eq10}) in terms of the gauge invariant
variables, ${^{(i)}}\!H_T$, ${^{(i)}}\!F$, ${^{(i)}}\!F_{ab}$ and
${^{(i)}}\!Z_a$. In order to do that, we use the expansion of $\triangle_L h_{\mu\nu}$  given in the appendix of \cite{seto} to simplify (\ref{eq10}).
For the sake of ready reference, we are giving the expressions borrowed from \cite{seto} here:
\begin{align}
2\triangle_L h_{ab}=&-\bar{D}^c\bar{D}_ch_{ab}+\bar{D}_b\bar{D}_ch^c_a+\bar{D}_a\bar{D}_ch^c_b+n\frac{\bar{D}^cr}{r}(-\bar{D}_ch_{ab}+\bar{D}_ah_{cb}+\bar{D}_bh_{ca})\nonumber\\&+{^{m}}\!R_a^ch_{cb}
+{^{m}}\!R^c_bh_{ca}-2{^{m}}\!R_{acbd}h^{cd}-\frac{1}{r^2}\hat{\triangle}h_{ab}+\frac{1}{r^2}(\bar{D}_a\bar{D}^ih_{bi}+\bar{D}_b\bar{D}^ih_{ai})
\nonumber\\&-\frac{\bar{D}_br}{r^3}\bar{D}_ah_{ij}\gamma^{ij}-\frac{\bar{D}_ar}{r^3}\bar{D}_bh_{ij}\gamma^{ij}+\frac{4}{r^4}\bar{D}_ar\bar{D}_brh_{ij}\gamma^{ij}-\bar{D}_a\bar{D}_bh
\label{lh10}
\end{align}
\begin{align}
2\triangle_L h_{ai}=&\bar{D}_i\bar{D}_bh^b_a+\frac{n-2}{r}\bar{D}^br\bar{D}_ih_{ab}-r\bar{D}^c\bar{D}_c\left(\frac{1}{r}h_{ai} \right)
\nonumber\\&-\frac{n}{r}\bar{D}^br\bar{D}_bh_{ai}-\bar{D}_ar\bar{D}_b\left(\frac{1}{r}h_i^b  \right)+\frac{n+1}{r}\bar{D}^br\bar{D}_ah_{bi}
\nonumber\\&+\left((n+1)\frac{(\bar{D}r)^2}{r^2}+(n-1)\frac{(1-\bar{D}r)^2}{r^2} -\frac{\bar{D}^c\bar{D}_cr}{r} \right)h_{ai}
\nonumber\\&  +r\bar{D}_a\bar{D}_b\left( \frac{1}{r}h_i^b \right) +\frac{1}{r^2}\bar{D}^br\bar{D}_arh_{bi}+(n+1)r\bar{D}_a\left( \frac{1}{r^2}\bar{D}^br \right)h_{bi}
\nonumber\\&-\frac{n+2}{r}\bar{D}_a\bar{D}^brh_{bi}+{^{m}}\!R_a^bh_{ib}-\frac{1}{r^2}\hat{\triangle}h_{ai}
\nonumber\\&+\frac{1}{r^2}\bar{D}_i\bar{D}^jh_{aj}
+r\bar{D}_a\left(\frac{1}{r^3}\bar{D}^jh_{ij} \right)+\frac{1}{r^3}\bar{D}_ar\bar{D}^jh_{ij}
\nonumber\\&-\frac{1}{r^3}\bar{D}_ar\bar{D}_ih_{jk}\gamma^{jk}
-r\bar{D}_a\left( \frac{1}{r}\bar{D}_ih \right)
\label{lh11}
\end{align}
\begin{align}
2\triangle_Lh_{ij}=&[2r\bar{D}^ar\bar{D}_bh_a^b+2(n-1)\bar{D}^ar\bar{D}^brh_{ab}+2r\bar{D}^a\bar{D}^brh_{ab}]\gamma_{ij}
\nonumber\\&+r\bar{D}_i\bar{D}_a\left( \frac{1}{r}h_j^a \right)+r\bar{D}_j\bar{D}_a\left(\frac{1}{r}h_i^a  \right)+(n-1)\frac{\bar{D}^ar}{r}(\bar{D}_ih_{aj}+\bar{D}_jh_{ai})
\nonumber\\&+2\frac{\bar{D}^ar}{r}\bar{D}^kh_{ka}\gamma_{ij}
-r^2\bar{D}^c\bar{D}_c\left(\frac{1}{r^2}h_{ij}\right)-n\frac{\bar{D}^ar}{r}\bar{D}_ah_{ij}
\nonumber\\&-\frac{1}{r^2}\hat{\triangle}h_{ij}+2\left( \frac{(n-1)}{r^2}+2\frac{(\bar{D}r)^2}{r^2} -\frac{\bar{D}^c\bar{D}_cr}{r}\right)h_{ij}
\nonumber\\&  +\frac{1}{r^2}(\bar{D}_i\bar{D}^kh_{kj}+\bar{D}_j\bar{D}^kh_{ki})
-2(\gamma^{kl}h_{kl}\gamma_{ij}-h_{ij})\frac{(1-(\bar{D}r)^2)}{r^2}
\nonumber\\&-2\frac{(\bar{D}r)^2}{r^2}\gamma_{ij}\gamma^{kl}h_{kl}-\bar{D}_i\bar{D}_jh
-r\bar{D}^ar\bar{D}_ah\gamma_{ij}
\label{lh12}
\end{align}
We then make the substitution (\ref{lh10} - \ref{lh12}) as well as (\ref{eq26}) in (\ref{eq10}) to get the various higher order equations (for $i=1$, we of course have ${^{(i)}}\!S_{\mu\nu}=0$ and get back the linearized equations presented in detail in \cite{seto}).

 For $\mu=i,\nu=j$, (\ref{eq10}) takes the form
\begin{align}
&\sum_{\textbf{k}}  \Bigg[-r^2\bar{D}^a\bar{D}_a{^{(i)}}\!H_T-nr\bar{D}^ar\bar{D}_a{^{(i)}}\!H_T+(k^2+2K){^{(i)}}\!H_T\Bigg]_{\textbf{k}}\mathbb{T}_{\textbf{k}ij}\nonumber\\&+\sum_{\textbf{k}_v} \Bigg[-\frac{2k_v}{r^{n-2}}\bar{D}_a(r^{n-1}{^{(i)}}\!Z^a)\Bigg]_{\textbf{k}_v}\mathbb{V}_{\textbf{k}_vij}+\sum_{\textbf{k}_s}\Bigg[-k_s^2[2(n-
2){^{(i)}}\!F+{^{(i)}}\!F_c^c]\Bigg]_{\textbf{k}_s}\mathbb{S}_{\textbf{k}_sij}\nonumber\\&={^{(i)}}\!S_{ij}-\sum_{\textbf{k}_s}[Q_4]_{\textbf{k}_s}\gamma_{ij}\mathbb{S}_{\textbf{k}_s}.
\label{lh1}
\end{align}
 We do not write the explicit form of $[Q_4]$  as it is not required in our
calculations and it does not contribute when we finally project to individual tensor components of each type.\\

Similarly  for $\mu=a,\nu=i$, (\ref{eq10}) takes the form,
\begin{align}
&\sum_{\textbf{k}_v}\Bigg[-\frac{1}{r^n}\bar{D}^b\left\{
r^{n+2}\left[\bar{D}_b\left(\frac{{^{(i)}}\!Z_a}{r}\right)-\bar{D}_a\left(\frac{{^{(i)}}\!Z_b}{r}\right)\right]\right\}+\frac{k^2_v-(n-1)K}{r}{^{(i)}}\!Z_a\Bigg]_{\textbf{k}_v}\mathbb{V}_{\textbf{k}_vi}\nonumber\\&+\sum_{\textbf{k}_s}\Bigg[-k_s\Big(\frac{1}{r^{n-2}}\bar{D}_b(r^{n-2}{^{(i)}}\!F^b_a)-r\bar{D}_a\bigg(\frac{1}{r}{^{(i)}}\!F^b_b\bigg)-2(n-1)\bar{D}_a{^{(i)}}\!F\Big)
\Bigg]_{\textbf{k}_s}\mathbb{S}_{\textbf{k}_si}\nonumber\\&={^{(i)}}\!S_{ai}.
\label{lh2}
\end{align}
In order to decompose the various sectors we use the fact that
\begin{align}
\int \mathbb{T}^{ij}\mathbb{V}_{ij}d^n\Omega=\int
\mathbb{T}^{ij}\mathbb{S}_{ij}d^n\Omega=\int
\mathbb{V}^{ij}\mathbb{S}_{ij}d^n\Omega=\int
\mathbb{V}^i\mathbb{S}_id^n\Omega=0.
\label{lh3}
\end{align}
After doing the necessary projections, one gets the following
equations:\\
We first project $\mathbb{T}^{ij}$ on (\ref{lh1}) to get the higher order tensor sector:
\begin{align}
-r^2\bar{D}^a\bar{D}_a{^{(i)}}\!H_T-nr\bar{D}^ar\bar{D}_a{^{(i)}}\!H_T+(k^2+2){^{(i)}}\!H_T=\int
\mathbb{T}^{ij}_{\bf{k}}\:{^{(i)}}\!S_{ij}d^n\Omega.
\label{lh4}
\end{align}
Similarly, carrying out the projection using $\mathbb{V}_{ij}$ and $\mathbb{V}_i$ on (\ref{lh1}) and (\ref{lh2}) respectively, we obtain the two equations pertaining to the vector sector:
\begin{align}
-\frac{1}{r^n}\bar{D}^b\left\{
r^{n+2}\left[\bar{D}_b\left(\frac{{^{(i)}}\!Z_a}{r}\right)-\bar{D}_a\left(\frac{{^{(i)}}\!Z_b}{r}\right)\right]\right\}&+\frac{k^2_v-(n-1)}{r}{^{(i)}}\!Z_a\nonumber\\&=\int
\mathbb{V}^i_{\bf{k}_v}{^{(i)}}\!S_{ai} d^n\Omega,
\label{lh5}
\end{align}
\begin{align}
-\frac{2k_v}{r^{n-2}}\bar{D}_a(r^{n-1}{^{(i)}}\!Z^a)=\int
\mathbb{V}^{ij}_{\bf{k}_v}{^{(i)}}\!S_{ij} d^n\Omega.
\label{lh6}
\end{align}
Finally, for the scalar sector, we will be using the following three equations:
\begin{align}
&-\bar{D}^c\bar{D}_c {^{(i)}}\!F_{ab}+\bar{D}_a\bar{D}_c
{^{(i)}}\!F^c_b+\bar{D}_b\bar{D}_c
{^{(i)}}\!F^c_a+n\frac{\bar{D}^cr}{r}(-\bar{D}_c {^{(i)}}\!F_{ab}+\bar{D}_a
{^{(i)}}\!F_{cb}+\bar{D}_b
{^{(i)}}\!F_{ca})\nonumber\\&+{^{m}}\!R_a^c{^{(i)}}\!F_{cb}+{^{m}}\!R^c_b{^{(i)}}\!F_{ca}-2{^{m}}\!R_{acbd}{^{(i)}}\!F^{cd}+\Bigg(\frac{k_s^2}{r^2}+\frac{2(n+1)}{L^2}\Bigg){^{(i)}}\!F_{ab}-\bar{D}_a\bar{D}_b
{^{(i)}}\!F^c_c\nonumber\\&-2n\left(   \bar{D}_a\bar{D}_b
{^{(i)}}\!F+\frac{1}{r}\bar{D}_a r\bar{D}_b{^{(i)}}\!F+\frac{1}{r}\bar{D}_b
r\bar{D}_a{^{(i)}}\!F\right)-\Bigg[ \bar{D}_c\bar{D}_d
{^{(i)}}\!F^{cd}\nonumber\\&+\frac{2n}{r}\bar{D}^cr\bar{D}^d
{^{(i)}}\!F_{cd}+\Big(-{^{m}}\!R_{cd}+\frac{2n}{r}\bar{D}^c\bar{D}^dr+\frac{n(n-1)}{r^2}\bar{D}^cr\bar{D}^d
r\Big) {^{(i)}}\!F_{cd}\nonumber\\&-2n\bar{D}^c\bar{D}_c
{^{(i)}}\!F-\frac{2n(n+1)}{r}\bar{D}^cr\bar{D}_c{^{(i)}}\!F+2(n-1)\frac{(k^2_s-n)}{r^2}{^{(i)}}\!F-\bar{D}^c\bar{D}_c
{^{(i)}}\!F^d_d\nonumber\\&-\frac{n}{r}\bar{D}^c r \bar{D}_c
{^{(i)}}\!F^d_d+\frac{k^2_s}{r^2}{^{(i)}}\!F^d_d\Bigg]g_{ab}=\int
\mathbb{S}_{\bf{k}_s}{^{(i)}}\!S_{ab} d^n\Omega,
\label{lh7}
\end{align}
\begin{align}
-k_s\Big(\frac{1}{r^{n-2}}\bar{D}_b(r^{n-2}{^{(i)}}\!F^b_a)-r\bar{D}_a\left(\frac{1}{r}{^{(i)}}\!F^b_b\right)&-2(n-1)\bar{D}_a
{^{(i)}}\!F\Big)\nonumber
\\&=\int \mathbb{S}_{\bf{k}_s}^i {^{(i)}}\!S_{ai} d^n\Omega,
\label{lh8}
\end{align}
\begin{align}
-k^2_s[2(n-2){^{(i)}}\!F+{^{(i)}}\!F^c_c]=\int
\mathbb{S}^{ij}_{\bf{k}_s}{^{(i)}}\!S_{ij} d^n\Omega.
\label{lh9}
\end{align}
As mentioned earlier, we will drop the superscript $(1)$ on metric
${^{(1)}}\!h_{\mu\nu}$, while considering the leading order perturbations.

\section{Tensor perturbations}
\subsection{At linear level}
Tensor perturbations are solely present in the $\delta g_{ij}$ component i. e. $h_{ij}=r^2H_{T\textbf{k}}\mathbb{T}_{\textbf{k}ij}$. Following \cite{wald}, we let $H_{T\textbf{k}}=r^{-n/2}\Phi_{T\textbf{k}}$.
Then the tensor perturbations at leading order are governed by:
\begin{align}
\ddot{\Phi}_T-f^2\Phi_T''-f'f\Phi_T'+\left(
\frac{n(n-2)}{4}\frac{f^2}{r^2}+\frac{n}{2r}f'f+\frac{l(l+n-1)}{r^2}
\right)\Phi_T=0
\label{eq35}
\end{align}
Substituting  the ansatz $\Phi_T=\cos(\omega t+b)\phi$ in (\ref{eq35}) we
get:
\begin{align}
\hat{L}\phi=\omega^2\phi
\label{eq36}
\end{align}
where $\hat{L}$ is given by
\begin{align}
\hat{L}=-f^2\partial_r^2-f'f\partial_r+\left(\frac{n(n-2)}{4}\frac{f^2}{r^2}+\frac{n}{2r}f'f+\frac{l(l+n-1)}{r^2}
\right)
\label{eq37}
\end{align}
Out of the two linearly independent solutions of (\ref{eq36}), we choose the one which has a regular fall off (of the form $\frac{1}{r^{\frac{n}{2}+1}}$) as $r\rightarrow\infty$
\begin{align}
\phi=e_{p,l}=d_{p,l}\frac{L^{\frac{1}{2}+\nu}r^{\frac{1}{2}+\sigma}}{(r^2+L^2)^{\frac{(\nu+\sigma+1)}{2}}}{_{2}}F_1\left(\zeta^{\omega}_{\nu,\sigma},\zeta_{\nu,\sigma}^{-\omega},1+\nu;\frac{L^2}{(r^2+L^2)}\right)
\label{eq38}
\end{align}
where $\zeta^{\omega}_{\nu,\sigma}=\frac{\nu+\sigma+\omega L+1}{2}$,
$\nu=\frac{(n+1)}{2}$
and $\sigma=l+\frac{(n-1)}{2}$
The eigenfrequencies $\omega$ are determined by imposing regularity of
$\phi$
at the origin, which gives us
\begin{align}
\omega L=2p+l+n+1;~~p=0,1,2...
\label{eq39}
\end{align}
The eigenfunctions $e_{p,l}$ form a complete orthogonal set  w.r.t the
inner product
\begin{align}
<e_{p,l},e_{p',l}>=\int_0^{\infty}e_{p,l}e_{p',l}w(r)dr=\delta_p^{p'}
\label{eq40}
\end{align}
where $w(r)$ is the appropriate weight function given by
\begin{align}
w(r)=\frac{1}{f}
\label{eq41}
\end{align}
Hence the normalization constant $d_{p,l}$ is given by
\begin{align}
d_{p,l}=\left[\frac{2}{L}\frac{(2p+l+n+1)\Gamma(p+l+n+1)}{p!\Gamma(p+l+\frac{n+1}{2})\Gamma(p+\frac{n+3}{2})}\right]^{1/2}\left(
\frac{n+3}{2} \right)_p
\label{eq42}
\end{align}

\subsection{Higher orders}
To study higher order tensor perturbations, we let
${^{(i)}}\!H_{T\textbf{k}}=r^{-\frac{n}{2}}{^{(i)}}\!\Phi_{T\textbf{k}}$ in (\ref{lh4}), which leads to
\begin{align}
{^{(i)}}\!\ddot{\Phi}_T+\hat{L}{^{(i)}}\!\Phi_T=r^{\frac{n}{2}-2}f\int
\mathbb{T}^{ij}{^{(i)}}\!S_{ij}d^n\Omega
\label{eq43}
\end{align}
The solution to the above equation can be written as
\begin{align}
{^{(i)}}\!\Phi_T={^{(i)}}\!\Phi_T^{(H)}+{^{(i)}}\!\Phi_T^{(P)}
\label{th1}
\end{align}
where ${^{(i)}}\!\Phi_T^{(H)}$ satisfies the homogeneous part of equation (\ref{eq43}). Clearly, as $r\rightarrow\infty$, its leading order behaviour is like $\frac{1}{r^{(n/2+1)}}$. The behaviour of $ {^{(i)}}\!\Phi_T^{(P)}$ part, depends on the source on the R.H.S. of equation (\ref{eq43}).
We assume the following ansatz for ${^{(i)}}\!\Phi_T^{(P)}$ in the limit $r\rightarrow\infty$:
\begin{align}
{^{(i)}}\!\Phi_T^{(P)}=\frac{{^{(i)}}\!a_k}{r^k}+\frac{{^{(i)}}\!a_{k+1}}{r^{k+1}}+...
\label{th2}
\end{align}
and plug this expansion back in (\ref{eq43}) to expand the equation in large $r$ limit . Since, the R.H.S. of (\ref{eq43}) has a leading order behaviour $\frac{1}{r^{3n/2}}$ as $r\rightarrow\infty$, one can see that the leading term in ${^{(i)}}\!\Phi_T^{(P)}$ will falls off like $\frac{1}{r^{\frac{3n}{2}+2}}$ in this limit. Hence, as $r\rightarrow\infty$, ${^{(i)}}\!\Phi_T$ behaves like
\begin{align}
{^{(i)}}\!\Phi_T\sim\frac{1}{r^{\frac{n}{2}+1}}+\mathcal{O}(r^{-(\frac{n}{2}+2)})
\label{eq44}
\end{align}
This automatically ensures the correct leading order asymptotic behaviour of the tensor sector of $\delta g_{ij}$,
\begin{align}
{^{(i)}}\!h_{ij}&=\sum_{\textbf{k}}r^{2}H_{T\textbf{k}}\mathbb{T}_{\textbf{k}ij}\nonumber\\&
=\sum_{\textbf{k}}r^{2-\frac{n}{2}}\Phi_{T\textbf{k}}\mathbb{T}_{\textbf{k}ij}\nonumber\\&
\sim \frac{1}{r^{n-1}}+\ldots \mbox{  as  }r\rightarrow\infty
\label{th3}
\end{align}
Thus, the tensor sector at all orders are already in aAdS form.

Further simplification of equation (\ref{eq43}) can be done by using the orthonormality and completeness of eigenfunctions. The completeness of $e_{p,l}$ allows one to write ${^{(i)}}\Phi_T$ as
${^{(i)}}\!\Phi_{T\textbf{k}}=\sum_{p=0}^{\infty}{^{(i)}}\!c_{p,\textbf{k}}(t)e_{p,l}(r)$, so that ${^{(i)}}\!c_{p,\textbf{k}}$ satisfies:
\begin{align}
{^{(i)}}\!\ddot{c}_{p,\textbf{k}}(t)+\omega^2{^{(i)}}\!c_{p,\textbf{k}}(t)=<
r^{\frac{n}{2}-2}f\int \mathbb{T}^{ij}_{\textbf{k}}{^{(i)}}\!S_{ij}d^n\Omega,e_{p,l} >
\label{eq45}
\end{align}

\section{Vector perturbations}
\subsection{Linear level}
The following two independent equations govern vector-type perturbations:
\begin{align}
\dot{Z}_t-f^2Z_r'-f'fZ_r-\frac{(n-1)f^2}{r}Z_r=0
\label{eq46}
\end{align}
\begin{align}
\frac{r}{f}\ddot{Z}_r-\frac{r}{f}\dot{Z}_t'+\frac{1}{f}\dot{Z}_t+\left(\frac{k_v^2-(n-1)}{r}
\right)Z_r=0
\label{eq47}
\end{align}
From the above equations, one can obtain the following master equation in variable  $\Phi_{v\textbf{k}_v}$, which is defined as  $Z_{r\textbf{k}_v}=f^{-1}r^{-\frac{(n-2)}{2}}\Phi_{v\textbf{k}_v}$:
\begin{align}
\ddot{\Phi}_v-f^2\Phi_v''-f'f\Phi_v'+\left(\frac{n(n+2)}{4}\frac{f^2}{r^2}+(l_v(l_v+n-1)-n)\frac{f}{r^2}-\frac{n}{2}\frac{f'f}{r}
\right)\Phi_v=0
\label{eq49}
\end{align}
By letting $\Phi_v=\cos(\omega t+b)\phi_v$, (\ref{eq49}) becomes
\begin{align}
\hat{L}_v\phi_v=\omega^2\phi_v
\label{eq50}
\end{align}
Here, $\hat{L}_v$ is given by
\begin{align}
\hat{L}_v=-f^2\partial_r^2-ff'\partial_r+\left(\frac{n(n+2)}{4}\frac{f^2}{r^2}+(l_v(l_v+n-1)-n)\frac{f}{r^2}-\frac{n}{2}\frac{ff'}{r}
\right)
\label{eq51}
\end{align}
The change of variables have been done in such a way so as to ensure that the operator $\hat{L}_v$ is the same as given in \cite{wald} for the case linearized vector perturbations. Out of the two linearly independent solutions, we choose the one which has a regular fall off (of the form $\frac{1}{r^{n/2}}$) as $r\rightarrow\infty$.
\begin{align}
\phi_v=e^{(v)}_{p,l_v}=d^{(v)}_{p,l_v}\frac{L^{1/2+\nu_v}r^{1/2+\sigma_v}}{(r^2+L^2)^{\frac{1}{2}(1+\nu_v+\sigma_v)}}{_{2}}F_1\left(\zeta^{\omega_v}_{\nu_v,\sigma_v},\zeta^{-\omega_v}_{\nu_v,\sigma_v},1+\nu_{v};\frac{L^2}{(r^2+L^2)}\right)
\label{eq52}
\end{align}
where $\sigma_v=l_v+\frac{(n-1)}{2}$, $\nu_v=\frac{(n-1)}{2}$ and
$\zeta_{\nu_v,\sigma_v}^{\omega_v}=\frac{ \nu_v+\sigma_v+\omega_v L+1 }{2}$.
Regularity of the  eigensolution at origin sets the eigenfrequencies to be
\begin{align}
\omega_{v}L=2p+l_v+n;~~p=0,1,2...
\label{eq5353}
\end{align}
The vector modes also form a complete orthogonal set with an inner product
\begin{align}
<e^{(v)}_{p,l_v},e^{(v)}_{p',l_v}>_v=\int_0^{\infty}e_{p,l_v}^{(v)}e^{(v)}_{p',l_v}w_v(r)dr=\delta_p^{p'}
\label{eq5454}
\end{align}
where the weight function $w_v(r)$ is given by
\begin{align}
w_v(r)=\frac{1}{f}
\label{eq5555}
\end{align}
Hence the normalization constant $d^{(v)}_{p,l_v}$ is fixed as
\begin{align}
d^{(v)}_{p,l_v}=\left[\frac{2}{L}\frac{(2p+l_v+n)\Gamma(p+l_v+n)}{p!\Gamma(p+l_v+\frac{n+1}{2})\Gamma(p+\frac{n+1}{2})}\right]^{1/2}\left(\frac{n+1}{2}\right)_p
\label{eq56}
\end{align}
\subsection{Higher orders}
The higher order vector equations will be given in terms of the following
defined quantities:
\begin{align}
{^{(i)}}\!V_{s1\textbf{k}_v}=\int \mathbb{V}^{ij}_{\textbf{k}_v}{^{(i)}}\!S_{ij}d^n\Omega
\label{eq53}
\end{align}
\begin{align}
{^{(i)}}\!V_{s2\textbf{k}_v}=\int\mathbb{V}^i_{\textbf{k}_v}{^{(i)}}\!S_{ir}d^n\Omega
\label{eq54}
\end{align}
\begin{align}
{^{(i)}}\!V_{s3\textbf{k}_v}=
\Bigg[{^{(i)}}\!V_{s2\textbf{k}_v}-\frac{{^{(i)}}\!V_{s1\textbf{k}_v}}{k_vr}+\frac{1}{2k_vf}(f{^{(i)}}\!V_{s1\textbf{k}_v})'
\Bigg]
\label{eq55}
\end{align}
Then  the relevant equations are:
\begin{align}
{^{(i)}}\!\dot{Z}_t=f^2{^{(i)}}\!Z_r'+f'f{^{(i)}}\!Z_r+(n-1)\frac{f^2}{r}{^{(i)}}\!Z_r+\frac{f{^{(i)}}\!V_{s1}}{2k_vr}
\label{eq57}
\end{align}
\begin{align}
\frac{r}{f}{^{(i)}}\!\ddot{Z}_r-\frac{r}{f}{^{(i)}}\!\dot{Z}_t'+\frac{1}{f}{^{(i)}}\!\dot{Z}_t+\frac{k_v^2-(n-1)}{r}
{^{(i)}}\!Z_r={^{(i)}}\!V_{s2}
\label{eq58}
\end{align}
To get the master equation, we substitute the expression for ${^{(i)}}\!\dot{Z}_t$ given by (\ref{eq57})
in (\ref{eq58}) gives the following:
\begin{align}
\frac{r}{f}{^{(i)}}\!\ddot{Z}_r-rf{^{(i)}}\!Z_r''&-\left(3rf'+(n-2)f\right){^{(i)}}\!Z_r'+\Big(-rf''-\frac{r}{f}(f')^2\nonumber\\&-(2n-3)f'+2(n-1)\frac{f}{r}+\frac{k_v^2-(n-1)}{r}
\Big){^{(i)}}\!Z_r={^{(i)}}\!V_{s3}
\label{eq59}
\end{align}
Letting
${^{(i)}}\!Z_{r\textbf{k}_v}=f^{-1}r^{-\frac{(n-2)}{2}}{^{(i)}}\!\Phi_{v\textbf{k}_v}$
in (\ref{eq59})
gives us a master equation for ${^{(i)}}\!\Phi_v$:
\begin{align}
{^{(i)}}\!\ddot{\Phi}_v+\hat{L}_v{^{(i)}}\!\Phi_v=r^{\frac{n}{2}-2}f^2{^{(i)}}\!V_{s3}
\label{eq60}
\end{align}

The solution to above equation can be written as
\begin{align}
{^{(i)}}\!\Phi_v={^{(i)}}\!\Phi_v^{(H)}+{^{(i)}}\!\Phi_v^{(P)}
\label{vh1}
\end{align}
As $r\rightarrow\infty$, ${^{(i)}}\!\Phi_v^{(H)}$ which satisfies the homogeneous part of equation (\ref{eq60}), has a leading order behaviour $\frac{1}{r^{n/2}}$. Making use of the fact that the R.H.S. of (\ref{eq60}) falls off like $\frac{1}{r^{\frac{3n}{2}-1}}$, one can deduce the behaviour of ${^{(i)}}\!\Phi_v^{(P)}$ by expanding it as
\begin{align}
{^{(i)}}\!\Phi_v^{(P)}=\frac{{^{(i)}}\!a_k}{r^k}+\frac{{^{(i)}}\!a_{k+1}}{r^{k+1}}...
\label{vh2}
\end{align}
and plugging it back in (\ref{eq60}).  It turn out as $r\rightarrow\infty$, the leading order term in  ${^{(i)}}\!\Phi_v^{(P)}$ is like $\frac{1}{r^{\frac{3n}{2}+1}}$. Hence, leading order the fall off behaviour of ${^{(i)}}\!\Phi_v$ is
\begin{align}
{^{(i)}}\!\Phi_v\sim\frac{1}{r^{n/2}}
\label{vh3}
\end{align}

Further, to construct asymptotically AdS solutions to all orders,
we consider the  class of metric perturbations (with
suitable gauge choice) where ${^{(i)}}\!H_T^{(v)}=0$). For such a class, ${^{(i)}}\!f_a$ in (\ref{eq26}) is simply: ${^{(i)}}\!f_a={^{(i)}}\!Z_a$. Hence the metric perturbations, along with their gauge transformations take the following form (summation over $\textbf{k}_v$ on the R.H.S. of the following equations is implied):
\begin{align}
{^{(i)}}\!h_{ri}=\left[r^{2-\frac{n}{2}}f^{-1}{^{(i)}}\!\Phi_{v}-r^2\bar{D}_r\left(\frac{{^{(i)}}\!M^{(v)}}{r}\right)\right]\mathbb{V}_i
\label{eq61}
\end{align}
\begin{align}
{^{(i)}}\!h_{ti}=\left[\int^t\left(\frac{f}{r^{n-2}}\left(
r^{\frac{n}{2}}{^{(i)}}\!\Phi_v
\right)'+\frac{f}{2k_v}{^{(i)}}\!V_{s1}\right)dt-r{^{(i)}}\!\dot{M}^{(v)}\right]\mathbb{V}_i
\label{eq62}
\end{align}
\begin{align}
{^{(i)}}\!h_{ij}=rk_v{^{(i)}}\!M^{(v)}\:\mathbb{V}_{ij}
\label{eq63}
\end{align}
 \underline{\textbf{Gauge choice for vector perturbations:}}

In order to define ${^{(i)}}\!M^{(v)}$ appropriately, we notice from
(\ref{eq63}) that since ${^{(i)}}\!h_{ij}$ needs to fall off like
$r^{-(n-1)}$, ${^{(i)}}\!M^{(v)}$ should have an expansion of the following
form as $r\rightarrow\infty$:
\begin{align}
{^{(i)}}\!M^{(v)}=\frac{{^{(i)}}\!m_{n}}{r^n}+\ldots
\label{eq123}
\end{align}
As discussed earlier, in the limit $r\rightarrow\infty$, ${^{(i)}}\!\Phi_v$ is of the form:
\begin{align}
{^{(i)}}\!\Phi_v=\frac{{^{(i)}}\!\phi_{\frac{n}{2}}}{r^{\frac{n}{2}}}+\frac{{^{(i)}}\!\phi_{\frac{n}{2}+2}}{r^{\frac{n}{2}+2}}+\ldots
\label{eqd1}
\end{align}
We expand (\ref{eq61}) in the large $r$ limit and substitute (\ref{eqd1}) in it to obtain
\begin{align}
{^{(i)}}\!h_{ri}=&\frac{L^2}{r^{\frac{n}{2}}}\left( 1+\frac{L^2}{r^2} \right)^{-1}\left[ \frac{{^{(i)}}\!\phi_{\frac{n}{2}}}{r^{\frac{n}{2}}}+ \mathcal{O}(r^{-(\frac{n}{2}+2)}) \right]-r^2\left[\frac{{^{(i)}}\!m_n}{r^{n+1}} +\ldots\right]'\nonumber\\&
=L^2\left[\frac{{^{(i)}}\!\phi_{\frac{n}{2}}}{r^n}+\mathcal{O}(r^{-(n+2)})\right]+(n+1)\left[ \frac{{^{(i)}}\!m_n}{r^n}+\ldots \right]
\label{eqd2}
\end{align}
If we choose
\begin{align}
{^{(i)}}\!m_n=-\frac{L^2}{(n+1)}{^{(i)}}\!\phi_{\frac{n}{2}}
\label{eqd3}
\end{align}
then it will ensure that ${^{(i)}}\!h_{ri}$ has the correct leading order behaviour of the  form $\frac{1}{r^{n+2}}$ as $r\rightarrow\infty$. Hence, we can choose ${^{(i)}}\!M^{(v)}$ to be:
\begin{align}
{^{(i)}}\!M^{(v)}=-\frac{L^2}{(n+1)}r^{-n/2}{^{(i)}}\!\Phi_v
\label{eq64}
\end{align}
The above expression is similar to that given by \cite{rost} for $n=2$ and is applicable for linearized perturbations (where ${^{(1)}}\!S_{\mu\nu}=0$) as well.
 One can see that the source dependent term in (\ref{eq62}) falls off like
$r^{-(2n-2)}$ and hence doesn't spoil the aAdS boundary condition for
${^{(i)}}\!h_{ti}$ for the given choice of ${^{(i)}}\!M^{(v)}$ (even in the
lowest possible $n=2$ case).\\
Finally, we further simplify (\ref{eq60}). Because of completeness of $e^{(v)}_{p,l_v}$, one can write,
${^{(i)}}\!\Phi_{v\textbf{k}_v}=\sum_{p=0}^{\infty}{^{(i)}}\!c_{p,\textbf{k}_v}(t)e^{(v)}_{p,l_v}(r)$
where $c_{p,\textbf{k}_v}$ satisfies,
\begin{align}
{^{(i)}}\!\ddot{c}_{p,\textbf{k}_v}(t)+\omega_v^2{^{(i)}}\!c_{p,\textbf{k}_v}(t)=<r^{\frac{n}{2}-2}f^2{^{(i)}}\!V_{s3\textbf{k}_v},
e_{p,l_v}^{(v)}  >_v
\label{eq65}
\end{align}

\section{Scalar perturbations}
\subsection{Linear level}
The following equations govern scalar perturbations \cite{seto}:
\begin{align}
F^c_c+2(n-2)F=0
\label{eq66}
\end{align}
\begin{align}
\frac{nf}{r}\dot{F}_{rr}+\frac{k_s^2}{r^2}F_{rt}-2n\dot{F}'+\frac{nf'}{f}\dot{F}-\frac{2n}{r}\dot{F}=0
\label{eq67}
\end{align}
\begin{align}
\frac{nf^2}{r}F_{rr}'&+\Big(\frac{k_s^2}{r^2}f+\frac{2n}{r}f'f+n(n-1)\frac{f^2}{r^2}\Big)F_{rr}-2nfF''\nonumber\\&-\Big(nf'+2n(n+1)\frac{f}{r}\Big)F'+\frac{2(n-1)(k_s^2-n)}{r^2}F=0
\label{eq68}
\end{align}
\begin{align}
\frac{2n}{rf}\dot{F}_{rt}-\frac{2n}{f^2}\ddot{F}&-\frac{nf'}{r}F_{rr}+\frac{nf'}{f}F'+\frac{2n(n-1)}{r}F'-\frac{n(n-1)f}{r^2}F_{rr}+\frac{n}{r}(F_t^t)'\nonumber\\&-\frac{k_s^2}{fr^2}F_t^t-\frac{2(n-1)(k_s^2-n)}{fr^2}F=0
\label{eq69}
\end{align}
\begin{align}
\frac{1}{f}\dot{F}_{rt}+(F_t^t)'-\frac{1}{r}F_t^t+2(n-1)F'-\frac{(n-1)}{r}F_r^r-\frac{f'}{2}F_{rr}+\frac{f'}{2f}F_t^t=0
\label{eq70}
\end{align}
We will use the ansatz similar to \cite{soda} to simplify these equations, defined as
\begin{align}
F_{rt}=\frac{2r}{f}(\dot{\Phi}_s+\dot{F})
\label{eq71}
\end{align}
Hence, we obtain a single master equation in
terms of the master variable $\Phi_s$
\begin{align}
\ddot{\Phi}_s-f^2\Phi_s''-\left(f'f +\frac{nf^2}{r}
\right)\Phi_s'+\left(-(n-1)\frac{f'f}{r}+\frac{k_s^2}{r^2}f
\right)\Phi_s=0
\label{eq72}
\end{align}
Let $\Phi_s=\cos(\omega_{s} t+b)\phi_s$, then (\ref{eq72}) is
given by:
\begin{align}
\hat{L}_s\phi_s=\omega_s^2\phi_s
\label{eq73}
\end{align}
Here, $\hat{L}_s$ is defined as
\begin{align}
\hat{L}_s=-f^2\partial_r^2-\left(f'f+\frac{nf^2}{r}
\right)\partial_r+\left(-(n-1)\frac{f'f}{r}+\frac{k_s^2}{r^2}f\right)
\label{eq74}
\end{align}
This operator is the same as in \cite{wald} for linearized scalar perturbations. Out of the two linearly independent equations, we choose the one with a regular fall off (of the form $\frac{1}{r^{n-1}}$ ) as $r\rightarrow\infty$.
\begin{align}
\phi_s=e^{(s)}_{p,l_s}=d^{(s)}_{p,l_s}\frac{L^{\frac{1}{2}+\nu_s}r^{
\frac{1}{2}+\sigma_s-\frac{n}{2}}}{(r^2+L^2)^{\frac{1}{2}(\nu_s+\sigma_s+1)}}{_{2}}F_1\left(\zeta^{\omega_s}_{\nu_s,\sigma_s}
,\zeta^{-\omega_s}_{\nu_s,\sigma_s},1
+\nu_{s} ;\frac{L^2}{(r^2+L^2)}\right)
\label{eq75}
\end{align}
where $\nu_s=\frac{(n-3)}{2}$, $\sigma_s=l_s+\frac{(n-1)}{2}$ and
$\zeta_{\nu_s,\sigma_s}^{\omega_s}=\frac{\nu_s+\sigma_s+\omega_sL+1 }{2}$.
The eigenfrequencies  $\omega_s$ are obtained by imposing the regularity
condition at the
origin, which gives
\begin{align}
\omega_{s}L=2p+l_s+n-1~~;p=0,1,2...
\label{eq76}
\end{align}
The associated eigenfunctions $e^{(s)}_{p,l_s}$ form a complete orthogonal
set and the inner product is given by
\begin{align}
<e^{(s)}_{p,l_s},e^{(s)}_{p',l_s}>_s=\int_0^{\infty}e^{(s)}_{p,l_s}e^{(s)}_{p',l_s}w_s(r)dr
\label{eq77}
\end{align}
where the weight function $w_s(r)$ is given by
\begin{align}
w_s(r)=\frac{r^{n}}{f}
\label{eq78}
\end{align}
Hence the normalization constant is given by
\begin{align}
d^{(s)}_{p,l_s}=
\left(\frac{2}{L}\frac{(2p+l_s+n-1)\Gamma(p+l_s+n-1)}{p!\Gamma(p+l_s+\frac{n+1}{2})\Gamma(p+\frac{n-1}{2})
} \right)^{1/2}\left(\frac{n-1}{2}\right)_p
\label{eq79}
\end{align}
\subsection{Higher orders}
Before considering higher order perturbations, we define the following
quantities:
\begin{align}
{^{(i)}}\!S_{s0\textbf{k}_s}=\int \mathbb{S}^{ij}_{\bf{k}_s}{^{(i)}}\!S_{ij}d^n\Omega,
\label{eq80}
\end{align}
\begin{align}
{^{(i)}}\!S_{s1\textbf{k}_s}=\int \mathbb{S}_{\bf{k}_s}{^{(i)}}\!S_{rt} d^n\Omega,
\label{eq81}
\end{align}
\begin{align}
{^{(i)}}\!S_{s2\textbf{k}_s}=\int\mathbb{S}_{\bf{k}_s}{^{(i)}}\!S_{tt}d^n\Omega,
\label{eq82}
\end{align}
\begin{align}
{^{(i)}}\!S_{s3\textbf{k}_s}=\int\mathbb{S}_{\textbf{k}_s}{^{(i)}}\!S_{rr}d^n\Omega
\label{eq83}
\end{align}
\begin{align}
{^{(i)}}\!S_{s4\textbf{k}_s}=\frac{1}{k_s}\int\mathbb{S}^i_{\textbf{k}_s}{^{(i)}}\!S_{ir}d^n\Omega
\label{eq84}
\end{align}
\begin{align}
{^{(i)}}\!S_{s5\textbf{k}_s}=\left(\frac{k_s^2}{nr}+2f'+\frac{(n-1)f}{r}\right)\int^t{^{(i)}}\!S_{s1}dt-\frac{{^{(i)}}\!S_{s2}}{f}+\frac{f^2}{r}\left(\frac{r}{f}\int^t
{^{(i)}}\!S_{s1}dt\right)',
\label{eq85}
\end{align}
\begin{align}
{^{(i)}}\!S_{s6\textbf{k}_s}=\frac{f^2}{2n}\Bigg\{{^{(i)}}\!S_{s3}-\frac{n}{r}{^{(i)}}\!S_{s4}-\left(1-\frac{n}{k_s^2}\right)\frac{{^{(i)}}\!S_{s0}}{r^2f}&-\left(\frac{(k^2_s-n)}{nfr}-\frac{f'}{2f^2}\right)\int^t
{^{(i)}}\!S_{s1}dt\nonumber\\&+\frac{1}{f}{^{(i)}}\!S_{s5}-\frac{1}{k_srf}\left(nr^2f{^{(i)}}\!S_{s5}\right)'\Bigg\}
\label{eq86}
\end{align}
Then the following equations govern scalar perturbations:
\begin{align}
-k_s^2[{^{(i)}}\!F^c_c+2(n-2){^{(i)}}\!F]={^{(i)}}\!S_{s0}
\label{eq87}
\end{align}
\begin{align}
\frac{n}{r}f{^{(i)}}\!\dot{F}_{rr}+\frac{k_s^2}{r^2}{^{(i)}}\!F_{rt}-2n{^{(i)}}\!\dot{F}'+n\frac{f'}{f}{^{(i)}}\!\dot{F}-\frac{2n}{r}{^{(i)}}\!\dot{F}={^{(i)}}\!S_{s1}
\label{eq88}
\end{align}
\begin{align}
\frac{nf^2}{r}{^{(i)}}\!F_{rr}'&+\Big(\frac{k_s^2}{r^2}f+\frac{2n}{r}f'f+n(n-1)\frac{f^2}{r^2}\Big){^{(i)}}\!F_{rr}-2nf{^{(i)}}\!F''\nonumber\\&-\Big(nf'+2n(n+1)\frac{f}{r}\Big){^{(i)}}\!F'+2(n-1)\frac{(k_s^2-n)}{r^2}{^{(i)}}\!F=\frac{{^{(i)}}\!S_{s2}}{f}
\label{eq89}
\end{align}
\begin{align}
\frac{2n}{rf}{^{(i)}}\!\dot{F}_{rt}&-\frac{2n}{f^2}{^{(i)}}\!\ddot{F}-\frac{nf'}{r}{^{(i)}}\!F_{rr}+\frac{nf'}{f}{^{(i)}}\!F'+\frac{2n(n-1)}{r}{^{(i)}}\!F'-\frac{n(n-1)}{r^2}f{^{(i)}}\!F_{rr}\nonumber\\&+\frac{n}{r}({^{(i)}}\!F_t^t)'-\frac{k_s^2}{fr^2}{^{(i)}}\!F_t^t-\frac{2(n-1)(k_s^2-n)}{fr^2}{^{(i)}}\!F={^{(i)}}\!S_{s3}
\label{eq90}
\end{align}
\begin{align}
\frac{1}{f}{^{(i)}}\!\dot{F}_{rt}+({^{(i)}}\!F_t^t)'-\frac{1}{r}{^{(i)}}\!F_t^t+2(n-1){^{(i)}}\!F'-\frac{(n-1)}{r}{^{(i)}}\!F_r^r&-\frac{f'}{2}{^{(i)}}\!F_{rr}+\frac{f'}{2f}{^{(i)}}\!F_t^t\nonumber\\&={^{(i)}}\!S_{s4}
\label{eq91}
\end{align}
Using the ansatz:
\begin{align}
{^{(i)}}\!F_{rt}=\frac{2r}{f}({^{(i)}}\!\dot{\Phi}_s+{^{(i)}}\!\dot{F})
\label{eq92}
\end{align}
it is possible to use the  system of five equations to obtain a single
equation in
terms of the higher order master variable ${^{(i)}}\!\Phi_s$
\begin{align}
{^{(i)}}\!\ddot{\Phi}_s+\hat{L}_s{^{(i)}}\!\Phi_s={^{(i)}}\!S_{s6}
\label{eq93}
\end{align}

Similar to tensor and vector modes, the solution to the above equation can be written as
\begin{align}
{^{(i)}}\!\Phi_s={^{(i)}}\!\Phi_s^{\mathcal{H}}+{^{(i)}}\!\Phi_s^{\mathcal{P}}
\label{eq94}
\end{align}
where the behaviour of $ {^{(i)}}\!\Phi_s^{\mathcal{H}}$ as $r\rightarrow\infty$ is similar to homogeneous solution of (\ref{eq93}), i.e.
\begin{align}
{^{(i)}}\!\Phi_s^{\mathcal{H}}=\frac{1}{r^{n-1}}\left({^{(i)}}\!a_{n-1}+\frac{
{^{(i)}}\!a_{n+1} }{r^2}+....\right)
\label{eq95}
\end{align}
The nature of the ${^{(i)}}\!\Phi_s^{\mathcal{P}}$ a  can be deduced by
looking at the behaviour of ${^{(i)}}\!S_{s6}$, whose leading order
behaviour as $r\rightarrow\infty$ goes like $r^{-(2n-2)}$. Hence, in this limit
\begin{align}
{^{(i)}}\!\Phi_s^{\mathcal{P}}=\frac{{^{(i)}}\!b_{2n}}{r^{2n}}+O(r^{-(2n+1)})
\label{eq96}
\end{align}

The various gauge invariant quantities in terms of ${^{(i)}}\!\Phi_s$ are
as follows.
\begin{align}
{^{(i)}}\!F=\frac{1}{(-k^2+n)}\left[nrf{^{(i)}}\!\Phi_s'+(k_s^2+n(n-1)f){^{(i)}}\!\Phi_s-\frac{nr^2f}{2k_s^2}{^{(i)}}\!S_{s5}
\right]
\label{eq97}
\end{align}
Rest of the variables can be expressed in terms of ${^{(i)}}\!F$ and
${^{(i)}}\!\Phi_s$. For e.g.
\begin{align}
{^{(i)}}\!F_{rr}=\frac{2r}{f}{^{(i)}}\!F'+\frac{(-k_s^2+n)}{nf^2}{^{(i)}}\!F-\frac{2k_s^2}{nf^2}{^{(i)}}\Phi_s+\frac{r}{nf}\int
{^{(i)}}\!S_{rt}dt
\label{eq98}
\end{align}
Similarly, ${^{(i)}}\!F_{tt}$ is obtained from (\ref{eq87}).
In order to construct aAdS solutions, we consider a class of perturbations where
${^{(i)}}\!H_L={^{(i)}}\!f_a=0$ at each order. For this choice, ${^{(i)}}\!f_{ab}={^{(i)}}\!F_{ab}$ and ${^{(i)}}H_L^{(s)}={^{(i)}}\!F$. Hence the metric
perturbations along with the gauge transformations  are given as (summation over $ \textbf{k}_s$ on the R.H.S. of each of the equations is implied):
\begin{align}
{^{(i)}}\!h_{tt}=\left[{^{(i)}}\!F_{tt}-2{^{(i)}}\!\dot{T}_t+f'f{^{(i)}}\!T_r\right]\mathbb{S}
\label{eq99}
\end{align}
\begin{align}
{^{(i)}}\!h_{rr}=\left[
{^{(i)}}\!F_{rr}-2{^{(i)}}\!T_r'-\frac{f'}{f}{^{(i)}}\!T_r \right]\mathbb{S}
\label{eq100}
\end{align}
\begin{align}
{^{(i)}}\!h_{rt}=\left[
{^{(i)}}\!F_{rt}-{^{(i)}}\!T_t'-{^{(i)}}\!\dot{T}_r+\frac{f'}{f}{^{(i)}}\!T_t\right]\mathbb{S}
\label{eq101}
\end{align}
\begin{align}
{^{(i)}}\!h_{ti}=\left[-{^{(i)}}\!\dot{M}+k_s{^{(i)}}\!T_t\right]\mathbb{S}_i
\label{eq102}
\end{align}
\begin{align}
{^{(i)}}\!h_{ri}=\left[- r^2\left( \frac{{^{(i)}}\!M}{r} \right)'+k_sT_r
\right]\mathbb{S}_i
\label{eq103}
\end{align}
\begin{align}
{^{(i)}}\!h_{ij}=2\left[
r^2{^{(i)}}\!F-\frac{k_sr}{n}{^{(i)}}\!M-rf{^{(i)}}\!T_r
\right]\gamma_{ij}\mathbb{S}+2k_sr\:{^{(i)}}\!M\mathbb{S}_{ij}
\label{eq104}
\end{align}
In order to ensure that the metric perturbations remain asymptotically AdS,
we need to make suitable gauge choices. \\
\underline{\textbf{Gauge choice for scalar perturbations:}}
In order to ensure that the metric perturbations (\ref{eq99}-\ref{eq104}) satisfy asymptotically AdS conditions we need to make
appropriate gauge choices. From the earlier discussion, we know ${^{(i)}}\!\Phi_s$ behaves in the following way as $r\rightarrow\infty$
\begin{align}
{^{(i)}}\!\Phi_s=\frac{{^{(i)}}\!\phi_{n-1}}{r^{n-1}}+\frac{{^{(i)}}\!\phi_{n+1}}{r^{n+1}}+\mathcal{O}(r^{-(n+3)})
\label{eqd4}
\end{align}
where the ellipsis denote the terms with lower powers of $r$.
Now we expand ${^{(i)}}\!F$ and ${^{(i)}}\!F_{rr}$ (as given by (\ref{eq97}) and (\ref{eq98}) respectively) in the large $r$ limit. They take the following form
\begin{align}
{^{(i)}}\!F=\frac{{^{(i)}}\!f_{n-1}}{r^{n-1}}+\frac{{^{(i)}}\!f_{n+1}}{r^{n+1}}+\ldots
\label{eqd5}
\end{align}
\begin{align}
{^{(i)}}\!F_{rr}=\frac{{^{(i)}}\!f^{(n+1)}_{rr}}{r^{n+1}}+\frac{{^{(i)}}\!f^{(n+3)}_{rr}}{r^{n+3}}+\ldots
\label{eqd6}
\end{align}
where
\begin{align}
{^{(i)}}\!f_{n-1}=\frac{1}{(-k_s^2+n)}\left (k_s^2{^{(i)}}\!\phi_{n-1}-\frac{2n}{L^2}{^{(i)}}\!\phi_{n+1} \right)
\label{eqd100}
\end{align}
and
\begin{align}
{^{(i)}}\!f^{(n+1)}_{rr}=-2L^2(n-1){^{(i)}}\!f_{n-1}
\label{eqd101}
\end{align}
We first assume that as $r\rightarrow\infty$,
${^{(i)}}\!T_r$ should have the following behaviour
\begin{align}
{^{(i)}}\!T_r=\frac{{^{(i)}}\!T_r^{(n)}}{r^n}+\ldots
\label{eq125}
\end{align}
Now
we put the expansions (\ref{eqd6}) and (\ref{eq125}) in (\ref{eq100}) and take the large $r$ limit to obtain
\begin{align}
{^{(i)}}\!h_{rr}=&\frac{{^{(i)}}\!f^{(n+1)}_{rr}}{r^{n+1}}+\frac{{^{(i)}}\!f^{(n+3)}_{rr}}{r^{n+3}}+\ldots-2\left[\frac{{^{(i)}}\!T_r^{(n)}}{r^n}+\ldots  \right]'\nonumber\\&-\frac{2}{r}\left( 1+\frac{r^2}{L^2}\right)^{-1}\left[\frac{{^{(i)}}\!T^{(n)}_r}{r^n}+\ldots \right]
\label{eqd7}
\end{align}
If we choose
\begin{align}
{^{(i)}}\!T^{(n)}_r=-\frac{1}{2(n-1)}{^{(i)}}\!f^{(n+1)}_{rr},
\label{eqd7.1}
\end{align}
then we can kill the terms which go like $\frac{1}{r^{n+1}}$, so that leading order behaviour of ${^{(i)}}\!h_{rr}$ is now the desired $\frac{1}{r^{n+3}}$ fall off.

Next, in order to ensure the correct aAdS behaviour of ${^{(i)}}\!h_{rt}$, we assume the following expansion for ${^{(i)}}\!T_t$ in the limit $r\rightarrow\infty$
\begin{align}
{^{(i)}}\!T_t=\frac{{^{(i)}}\!T_t^{(n-1)}}{r^{n-1}}+\ldots
\label{eq126}
\end{align}
Then we expand ${^{(i)}}\!h_{rt}$ (as given by (\ref{eq101})) in the large $r$ limit and substitute (\ref{eq126}) in it:

\begin{align}
{^{(i)}}\!h_{rt}=&2L^2\left( 1+\frac{L^2}{r^2}\right)^{-1}\left [ \frac{{^{(i)}}\!\phi_{n-1}}{r^{n}}+\frac{{^{(i)}}\!\dot{f}_{n-1}}{r^n}+\ldots \right]-\left[ \frac{{^{(i)}}\!T_r^{(n)}}{r^n}+\ldots\right]
\nonumber\\&-\left[ \frac{{^{(i)}}\!T_t^{(n-1)}}{r^{n-1}}  +\ldots \right]' +2\left( 1+\frac{L^2}{r^2} \right)^{-1}\left[ \frac{{^{(i)}}\!T_t^{(n-1)}}{r^n}+\ldots \right]
\label{eqd8}
\end{align}
By choosing
\begin{align}
{^{(i)}}\!T^{(n-1)}_t=\frac{1}{(n+1)}\left [ {^{(i)}}\!\dot{T}^{(n)}_r-2L^2{^{(i)}}\!\dot{f}_{n-1}-2L^2{^{(i)}}\!\dot{\phi}_{n-1} \right]
\label{eqd9}
\end{align}
we can kill off all the terms with $r^{-n}$ fall off, so that ${^{(i)}}\!h_{rt}$ has correct fall off.
Lastly, for ${^{(i)}}\!M$, we assume an expansion of the form
\begin{align}
{^{(i)}}\!M=\frac{{^{(i)}}\!m^{(s)}_n}{r^n}+\ldots
\label{eqd10}
\end{align}
and then substitute this in the expansion of (\ref{eq103}) in the large $r$ limit to obtain
\begin{align}
{^{(i)}}\!h_{ri}=-r^2\left[ \frac{{^{(i)}}\!m^{(s)}_n}{r^{n+1}}+\ldots  \right]'+k_s\left [\frac{{^{(i)}}\!T^{(n)}_r}{r^n}+\ldots \right]
\label{eqd11}
\end{align}
Choosing
\begin{align}
{^{(i)}}\!m^{(s)}_n=-\frac{k_s}{(n+1)}{^{(i)}}\!T_r^{(n)}
\label{eqd12}
\end{align}
will put (\ref{eq103}) in aAdS form.
Hence taking cue from (\ref{eqd7.1}), (\ref{eqd9}) and (\ref{eqd12}), we make the following constructions for ${^{(i)}}\!T_r$, ${^{(i)}}\!T_t$ and ${^{(i)}}\!M$:

\begin{align}
{^{(i)}}\!T_r=\frac{1}{(-k^2_s+n)}\left[\frac{k_s^2L^2}{r}{^{(i)}}\!\Phi_s+\frac{n}{r^{n-3}}\partial_r(r^{n-1}{^{(i)}}\!\Phi_s)
\right]
\label{eq105}
\end{align}
\begin{align}
{^{(i)}}\!T_t=-\frac{L^2}{(n+1)(-k_s^2+n)}\left[(-k_s^2+2n){^{(i)}}\!\dot{\Phi}_s+\frac{n}{L^2
r^{n-4}}\partial_r(r^{n-1}{^{(i)}}\!\dot{\Phi}_s)\right]
\label{eq106}
\end{align}
\begin{align}
{^{(i)}}\!M=-\frac{k_s}{(n+1)}{^{(i)}}\!T_r
\label{eq107}
\end{align}
These gauge choices are also valid for linearized perturbations $({^{(i)}}\!S_{\mu\nu}=0)$.

Finally, since $e^{(s)}_{p,l_s}$ form a complete orthonormal set, one can write
${^{(i)}}\!\Phi_s$ as
${^{(i)}}\!\Phi_{s\textbf{k}_s}=\sum_{p=0}^{\infty}{^{(i)}}\!c_{p,\textbf{k}_s}(t)e_{p,l_s}^{(s)}(r)$. From (\ref{eq93}), we see ${^{(i)}}\!c_{p,\textbf{k}_s}(t)$ satisfies:
\begin{align}
{^{(i)}}\!\ddot{c}_{p,\textbf{k}_s}+\omega_s^2{^{(i)}}\!c_{p,\textbf{k}_s}=<{^{(i)}}\!S_{s6\textbf{k}_s},e^{(s)}_{p,l_s}>_s
\label{eq108}
\end{align}

\section{Special modes}
These modes satisfy first order equations and beyond the linearized level,
are no longer gauge degrees of freedom.

\subsection{Scalar perturbations $l_s=0,1$ modes}
\subsubsection{$l_s=0$ mode}
Now we consider the $l_s=0$ mode for scalar perturbations. Let
${^{(i)}}\!\tilde{S}_{0\:\mu\nu}$ be the source terms associated with these
modes. For this case
$\mathbb{S}$ is just a constant and only  ${^{(i)}}\!f_{ab}$  and
${^{(i)}}\!H_L$ exist. We will make a gauge choice so that
\begin{align}
{^{(i)}}\!H_L ={^{(i)}}\!f_{rt}=0
\label{eq109}
\end{align}
We get the following equations for the
case ${^{(i)}}\!G_{rt}=0$, ${^{(i)}}\!G_{tt}=0$ and ${^{(i)}}\!G_{rr}=0$
respectively.
\begin{align}
\frac{nf}{r}{^{(i)}}\!\dot{f}_{rr}={^{(i)}}\!\tilde{S}_{0\:rt}
\label{eq110}
\end{align}
\begin{align}
\frac{nf}{r}({^{(i)}}\!f_r^r)'+\left(\frac{n(n-1)f}{r^2}+\frac{nf'}{r}
\right){^{(i)}}\!f_r^r=\frac{1}{f}{^{(i)}}\!\tilde{S}_{0\:tt}
\label{eq111}
\end{align}
\begin{align}
\frac{n}{r}({^{(i)}}\!f_t^t)'-\left(\frac{nf'}{fr}+\frac{n(n-1)}{r^2}\right){^{(i)}}\!f_{r}^r={^{(i)}}\!\tilde{S}_{0\:rr}
\label{eq112}
\end{align}
Upon solving (\ref{eq110}), one obtains
\begin{align}
{^{(i)}}\!f_{rr}=\int^t_{t_1} \frac{r}{nf}{^{(i)}}\!\tilde{S}_{0\:rt}
dt+{^{(i)}}\!f_{rr}(t_1,r)
\label{eq113}
\end{align}
  where ${^{(i)}}\!f_{rr}(t_1,r)$ can be obtained from (\ref{eq111})):
\begin{align}
{^{(i)}}\!f_{rr}(t_1,r)=\frac{1}{f^2r^{n-1}}\int_0^r\frac{r^n}{nf}{^{(i)}}\!\tilde{S}_{0\:tt}(t_1,r)dr
\end{align}
 Similarly ${^{(i)}}\!f_{tt}$ is given by
\begin{align}
{^{(i)}}\!f_{tt}=f^2{^{(i)}}\!f_{rr}-\frac{f}{n}\int^r_0 dr\left(
{^{(i)}}\!\tilde{S}_{0\:rr}+\frac{1}{f^2}{^{(i)}}\!\tilde{S}_{0\:tt}\right)r
\label{eq114}
\end{align}
\subsection{$l_s=1$ mode}
Let ${^{(i)}}\!\tilde{S}_{1\:\mu\nu}=\int
\mathbb{S}{^{(i)}}\!S_{\mu\nu}d^n\Omega$ be the source associated for this
mode.
Using gauge choice freedom,
${^{(i)}}\!H_L$ and ${^{(i)}}\!f_a^{(s)}$ is put to zero.
So we  need to solve for ${^{(i)}}\!f_{ab}$. From the ${^{(i)}}\!G_{tt}$
equation
\begin{align}
{^{(i)}}\!f_{rr}'+\left(\frac{1}{rf}+\frac{2f'}{f}+\frac{(n-1)}{r}\right){^{(i)}}\!f_{rr}=\frac{r}{nf^3}{^{(i)}}\!\tilde{S}_{1\:tt}
\label{eq115}
\end{align}
Hence,
\begin{align}
{^{(i)}}\!f_{rr}=\frac{1}{r^nf^{3/2}}\left(\int^r_0
\frac{r^{n+1}}{nf^{3/2}}{^{(i)}}\!\tilde{S}_{1\:tt}dr\right)
\label{eq116}
\end{align}
From ${^{(i)}}\!G_{rt}=0$ and ${^{(i)}}\!G_{it}=0$, we obtain
\begin{align}
\frac{nf}{r}{^{(i)}}\!\dot{f}_{rr}+\frac{n}{r^2}{^{(i)}}\!f_{rt}={^{(i)}}\!\tilde{S}_{1\:rt}
\label{eq117}
\end{align}
\begin{align}
-f\sqrt{n}\left[{^{(i)}}\!f_{rt}'+\left(\frac{(n-2)}{r}
 +\frac{f'}{f}\right){^{(i)}}\!f_{rt}-{^{(i)}}\!\dot{f}_{rr}
\right]={^{(i)}}\!\tilde{S}_{1\:it}
\label{eq118}
\end{align}
Hence, from the two equations,
\begin{align}
{^{(i)}}\!f_{rt}=\frac{1}{f^{1/2}r^{n-1}}\left\{\int_0^r
\frac{r^{n-1}}{f^{1/2}}\left(\frac{r}{n}{^{(i)}}\!\tilde{S}_{1\:rt}-\frac{1}{\sqrt{n}}{^{(i)}}\!\tilde{S}_{1\:it}\right)dr\right\}
\label{eq119}
\end{align}
Similarly from ${^{(i)}}\!G_{ir}=0$ equation we get
\begin{align}
{^{(i)}}\!f_{tt}=rf^{1/2}\left\{\int^r f^{1/2} \left(
\frac{{^{(i)}}\!\dot{f}_{tr}}{rf}-\frac{(n-1)f}{r^2}{^{(i)}}\!f_{rr}-\frac{{^{(i)}}\!\tilde{S}_{ir}}{\sqrt{n}r}-\frac{f'}{2r}{^{(i)}}\!f_{rr}
\right)dr\right\}
\label{eq120}
\end{align}

\subsection{Vector modes $l_v=1$ mode}
Since $\mathbb{V}_{ij}$ is undefined, only ${^{(i)}}\!f_a$ exist. We will
use gauge freedom to put ${^{(i)}}\!f_t$ to zero. Let
${^{(i)}}\!\tilde{S}^{(v)}_{1\:ia}=\int \mathbb{S}^i{^{(i)}}\!S_{ia}
d^n\Omega$. Then from ${^{(i)}}\!G_{ir}=0$, one obtains
\begin{align}
{^{(i)}}\!\dot{f}_r=\frac{f}{r}\int_{t_1}^t{^{(i)}}\!\tilde{S}^{(v)}_{1\:ir}dt+{^{(i)}}\!\dot{f}_r(t_1,r)
\label{eq121}
\end{align}
where ${^{(i)}}\!\dot{f}_r(t_1,r)$ can be determined from
${^{(i)}}\!G_{it}=0$:
\begin{align}
{^{(i)}}\!\dot{f}_r(t_1,r)=\frac{1}{r^{n+1}}\int_0^r\frac{r^n}{f}{^{(i)}}\!\tilde{S}^{(v)}_{1\:it}(t_1,r)dr
\label{eq122}
\end{align}

\section{Second order Analysis}

In the case of pure gravitational perturbations in four dimensions, it is seen that for specific examples of single mode data,
resonances are completely absent at second order \cite{diasantos}. One key reason why it happens is that given single mode initial data, with frequency $\tilde{\omega}$, there are two kinds of frequencies  excited at the second order, namely $\omega=\{0,2\tilde{\omega}\}$. $\omega=2\tilde{\omega}$ is always even. In four dimensions, one has only scalar (or polar) and vector (or axial) sectors. Each harmonic is labelled by an azimuthal quantum number $m$ and polar quantum number $l$. For the examples considered in \cite{diasantos}, as far as the scalar sector is concerned, for single mode data (which can be a scalar or a vector seed), the polar quantum numbers $l$, which get excited, are always even numbers. Now, the scalar frequency spectrum in four dimensions is given by
\begin{align}
\omega=2p+l+1;~~p=0,1,2...
\end{align}
If $l$ is a even number, then it means that any possible resonant frequency can only be odd. But it means there can be no resonant frequency corresponding to $2\tilde{\omega}$.

A similar reasoning can be applied to the vector sector as well in four dimensions, where no matter what initial data is taken, the excited harmonics always correspond to odd $l$. But since, the vector eigenfrequencies are denoted by $\omega=2p+l+2$, it means, having a odd $l$ would mean a odd resonant frequency. But such a frequency can never be equal to $2\tilde{\omega}$.

So far, in most of the literature concerning resonant instability in weakly nonlinear perturbation theory, resonances enter only at third order. Some of these resonances are irremovable (for a proposal to construct geons starting from any linear eigenfrequency, see \cite{rostcomment}). The characteristic feature of such irremovable resonances is that their quantum numbers are not the same as that of the initial seed. Hence, we cannot perform frequency shifts to remove such secular terms.

We now move on to dimension five, which is equivalent to $n=3$.
As an exercise, we consider a case, where the initial single mode data, contains only
tensor-type harmonics, with labels, $\textbf{k}=\{ l,l^{(1)},m \}$. Let the
single mode data have value
\begin{align}
\tilde{\textbf{k}}=\{l=2, l^{(1)}=2,m=0)
\label{eq124.11}
\end{align}
The associated source frequency spectrum is then $\tilde{\omega}L=2p'+6$ (see (\ref{eq39}).
Note that, even if we start out with a single tensor mode at the linear level, at higher orders, both scalar and vector modes could get excited as well.
For now though, we will restrict ourselves to the tensor sector. The higher order tensor sector equation is governed by (\ref{eq45}).
It is possible that even just in the tensor sector, such a initial seed could excite many kinds of tensor harmonics. Let us try to check the presence of a tensor harmonic which has the same quantum numbers as the source frequency. In other words, we will be taking the projection of (\ref{eq45}) on a tensor harmonic $\mathbb{T}^{ij}_{\textbf{k}}$ with $\textbf{k}=\tilde{\textbf{k}}$.
Let us now inspect the R.H.S of (\ref{eq45}) when $i=2$:
\begin{align}
<r^{-1/2}f\int \mathbb{T}^{ij}_{\tilde{\textbf{k}}}\;
{^{(2)}}\!S_{ij} d^3\Omega, e_{p,2}>=<r^{-1/2}f\int
\mathbb{T}^{ij}_{\tilde{\textbf{k}}}\;{^{(2)}}\!A_{ij} d^3\Omega,
e_{p,2}>
\label{eq125.11}
\end{align}
The above replacement of ${^{(2)}}\!S_{ij}$ with ${^{(2)}}\!A_{ij}$ is
possible because of the traceless property of the tensor harmonics, $\mathbb{T}^i_i=0$.
The general explicit form of ${^{(2)}}\!A_{ij}$, in case we start out only with
tensor type perturbations at linear level, can be obtained by writing ${^{(2)}}\!A_{ij}$ (see (\ref{eq9})), in terms of $\bar{D}_a$ and $\bar{D}_i$ and making the substitution $h_{ij}=r^2H_{T\textbf{k}}\mathbb{T}_{\textbf{k}ij}$:
\begin{align}
{^{(2)}}\!A_{ij}=&\sum_{\bf{k_1}}  \sum_{\bf{k_2}}H_{T\bf{k_1}}H_{T\bf{k_2}}\Big(\mathbb{T}^{kl}_{\bf{k_1}}(-\bar{D}_i\bar{D}_j\mathbb{T}_{kl_{\bf{k_2}}}+\bar{D}_k\bar{D}_i\mathbb{T}_{jl_{\bf{k_2}}}+\bar{D}_k\bar{D}_j\mathbb{T}_{li_{\bf{k_2}}}
\nonumber\\&-\bar{D}_k\bar{D}_l\mathbb{T}_{ij_{\bf{k_2}}})-\frac{\bar{D}_i\mathbb{T}^{kl}_{\bf{k_1}}\bar{D}_j\mathbb{T}_{kl_{\bf{k_2}}}}{2}+\bar{D}_k\mathbb{T}^l_{i_{\bf{k_1}}}\bar{D}_l\mathbb{T}^k_{j_{\bf{k_2}}}-\bar{D}^k\mathbb{T}_{il_{\bf{k_1}}}\bar{D}_k\mathbb{T}^l_{j_{\bf{k_2}}}\Big)\nonumber\\&
-r\bar{D}^ar\bar{D}_aH_{T\bf{k_1}}H_{T\bf{k_2}}\gamma_{ij} \mathbb{T}^{kl}_{\bf{k_1}}\mathbb{T}_{kl_{\bf{k_2}}}-r^2\bar{D}^aH_{T\bf{k_1}}\bar{D}_aH_{T\bf{k_2}}\mathbb{T}_{ik_{\bf{k_1}}}\mathbb{T}^k_{j_{\bf{k_2}}}
\label{eqapp1}
\end{align}
For single mode initial data, we have $\textbf{k}_1=\textbf{k}_2=\tilde{\textbf{k}}$ in (\ref{eqapp1}), and hence there won't be any summation over various $\textbf{k}$ indices. Let us first tackle the kind of integral, arising as a result of a term like
$H_{\textbf{k}_1}H_{\textbf{k}_2}$ in ${^{(2)}}\!A_{ij}$. Of course, if the angular integrals accompanying this term vanish, then this term won't contribute. Suppose angular integrals don't vanish, and a tensor harmonic with index $\tilde{\textbf{k}}$ is indeed excited, then we show that the $t-r$ integral does vanish.
The integral of interest is
\begin{align}
\int_0^{\infty} r^{-1/2} (H_{T_{p',l=2}})^2 e_{p,2} dr\sim &\int_{-1}^{1}(1-y)^2 [
P^{2,3}_{p'}(y)]^2 \frac{d^p}{dy^p}[(1-y)^{p+2}(1+y)^{p+3}
]dy\nonumber\\&\times\:\cos^2\left(\tilde{\omega}t  \right)
\label{eq126.11}
\end{align}
where $r^2=L^2\frac{(1+y)}{(1-y)}$
 and we have used the fact that the hypergeometric function
${_2}F_1(p+\alpha+\beta+1 ,-p, 1+\alpha; z)\sim P^{\alpha,\beta}_p(1-2z)$ (the Jacobi polynomial of degree $p$). Hence,
\begin{align}
H_{T\tilde{\textbf{k}},p}(y)\sim(1-y)^2(1+y)P^{(2,3)}_p(y)\cos \tilde{\omega}t
\label{soa5}
\end{align}
\begin{align}
e_{2p+3,2}(y)\sim (1+y)^{7/4}(1-y)^{5/4}P^{(2,3)}_{2p+3}(y)
\label{soa6}
\end{align}
and the Jacobi polynomial in (\ref{soa6}) is rewritten using the formula
\begin{align}
P_p^{(\alpha,\beta)}(y)\sim  (1-y)^{-\alpha}(1+y)^{-\beta}\frac{d^p}{dy^p}\left[(1-y)^{p+\alpha}(1+y)^{p+\beta}\right]
\label{soa30}
\end{align}
As discussed before, the only frequency excited at the second order is  $2\tilde{\omega}$.
The resonant modes correspond to those frequencies $\omega$, which satisfy
$\omega_{p,l=2}=2\tilde{\omega}$. This translates to
\begin{align}
p=2p'+3
\label{eq128.11}
\end{align}
Hence, upon inspecting (\ref{eq126.11}), and doing integration by parts, one
can conclude that this particular integral vanishes. (Note that, the integration gives rise to boundary terms which will vanish. Moreover, the resultant integral that is left is also zero, because the integrand has a derivative operator acting $p=2p'+3$ times on
$(1-y)^2 [P^{2,3}_{p'}]^2$, which is a polynomial of degree $2p'+2$). This same method was used by \cite{adsbi5} to prove that the interaction coefficients at second order are zero.

Now let us focus on the second kind of integral which arises because  of the term $ -r\bar{D}^ar\bar{D}_aH_{T\textbf{k}_1}H_{T\textbf{k}_2}\gamma_{ij}\mathbb{T}^{kl}_{\textbf{k}_1}\mathbb{T}_{\textbf{k}_2kl}$
in ${^{(2)}}\!A_{ij}$. Since, $\mathbb{T}^i_{\:i}=0$, the part of the integral (\ref{eq125.11}), arising
out of a term proportional to $\gamma_{ij}$ in ${^{(2)}}\!A_{ij}$ also
vanishes.

The last part in integral (\ref{eq125.11}) arises out of term $r^2
\bar{D}^aH_{T\textbf{k}_1}\bar{D}_aH_{T\textbf{k}_2}$. Equation (\ref{eq45}) now becomes
\begin{align}
{^{(2)}}\!\ddot{c}_{p,\tilde{\textbf{k}}}+\omega^2_{p,l=2}{^{(2)}}\!c_{p,\tilde{\textbf{k}}}=\int_{S^3}\mathbb{T}^{ij}_{\tilde{\textbf{k}}}\mathbb{T}_{\tilde{\textbf{k}}ik}\mathbb{T}^k_{\tilde{\textbf{k}}j}d^3\Omega\:\left< r^{3/2}f\bar{D}^aH_{T\tilde{\textbf{k}},p}\bar{D}_aH_{T\tilde{\textbf{k}},p},e_{2p+3,2}(r)\right>
\label{soa3}
\end{align}
The angular integral can be performed exactly (the explicit forms of the tensor harmonics have been derived in the Appendix):
\begin{align}
\int_{S^3}\mathbb{T}^{ij}_{\tilde{\textbf{k}}}\mathbb{T}_{ik\tilde{\textbf{k}}}\mathbb{T}^k_{j\tilde{\textbf{k}}}d^3\Omega=-\frac{379}{210\sqrt{6}\pi}
\label{soa2}
\end{align}
This indicates that a harmonic corresponding to $\tilde{\textbf{k}}$ indeed gets excited.
Next, we turn our attention to the $t-r$ integral:
\begin{align}
&\int_0^{\infty}r^{3/2}e_{2p+3,2}\left(\frac{1}{f}(\partial_t H_{T\tilde{\textbf{k}},p})^2 -f(\partial_r H_{T\tilde{\textbf{k}},p})^2 \right) dr\nonumber\\&
=a_1\int_{-1}^1 (1+y)^3(1-y)P^{(2,3)}_{2p+3}(y)\Big[(p+3)^2(1+y)(1-y)^3(P^{(2,3)}_{p})^2\sin^2 \tilde{\omega}t\nonumber\\&\hspace{0.5 cm}-\left(\partial_y[(1-y)^2(1+y)P^{(2,3)}_p]\right)^2\cos^2\tilde{\omega}t\Big]dy\nonumber\\&
=a_1\int_{-1}^1 (1+y)^3(1-y) P^{(2,3)}_{2p+3}(y)\Big[ (p+3)^2(1+y)(1-y)^3(P^{(2,3)}_p)^2\sin^2(\tilde{\omega}t)\nonumber\\&\hspace{0.5 cm}-B(y)\cos^2(\tilde{\omega}t) \Big]dy\nonumber\\&
=a\sin^2\tilde{\omega}t-b\cos^2\tilde{\omega}t
\label{soa1}
\end{align}
where $a_1$ denotes the common constant factors in the integral and
\begin{align}
a=a_1\int_{-1}^1 (p+3)^2(1+y)^4(1-y)^4P^{(2,3)}_{2p+3}(y)(P^{(2,3)}_{p})^2
\label{soa7}
\end{align}
\begin{align}
b=a_1\int_{-1}^1 (1+y)^3(1-y)P^{(2,3)}_{2p+3}(y)B(y) dy
\label{soa8}
\end{align}
with $B(y)=(\partial_y[(1-y)^2(1+y)P^{(2,3)}_p])^2$. Using the various relations for Jacobi polynomials, namely,
\begin{align}
\frac{d}{dy}P_p^{(\alpha,\beta)}(y)=\frac{1}{2}(p+\alpha+\beta+1)P_{p-1}^{(\alpha+1,\beta+1)}
\label{soa22}
\end{align}
\begin{align}
P^{(\alpha,\beta-1)}_{p}(y)-P^{(\alpha-1,\beta)}_p(y)=P^{(\alpha,\beta)}_{p-1}(y)
\label{soa23}
\end{align}
\begin{align}
(1-y)P_p^{(\alpha+1,\beta)}(y)+(1+y)P_p^{(\alpha,\beta+1)}(y)=2P_p^{(\alpha,\beta)}(y)
\label{soa24}
\end{align}
\begin{align}
(2p+\alpha+\beta+1)P_p^{(\alpha,\beta)}(y)=(p+\alpha+\beta+1)P_p^{(\alpha,\beta+1)}(y)+(p+\alpha)P_{p-1}^{(\alpha,\beta+1)}(y)
\label{soa25}
\end{align}
we will now simplify $B(y)$.
\begin{align}
B(y)&=\left\{[-2(1-y^2)+(1-y)^2]P^{(2,3)}_p+(1+y)(1-y)^2\frac{d}{dy}P^{(2,3)}_{p} \right\}^2\nonumber\\&
=\left\{ [-2(1-y^2)+(1-y)^2]P^{(2,3)}_p+\frac{1}{2}(p+6)(1+y)(1-y)^2P_{p-1}^{(3,4)}\right\}\nonumber\\&
=\left\{ [-2(1-y^2)+(1-y)^2]P^{(2,3)}_p+\frac{1}{2}(p+6)(1+y)(1-y)^2[ P_{p}^{(3,3)} -P_p^{(2,4)}] \right\}^2\nonumber\\&
=\Big\{ [-2(1-y^2)+(1-y)^2]P^{(2,3)}_p+\frac{1}{2}(p+6)(1-y^2)[ (1-y)P_{p}^{(3,3)}\nonumber\\&\hspace{0.5 cm}-(1-y)P^{(2,4)}_p  ]  \Big\}^2\nonumber\\&
=\Big\{  [-2(1-y^2)+(1-y)^2]P^{(2,3)}_p+\frac{1}{2}(p+6)(1-y^2)[2P^{(2,3)}_p\nonumber\\&\hspace{0.5 cm}-(1+y)P_p^{(2,4)}-(1-y)P_p^{(2,4)}]\Big\}^2\nonumber\\&
=\left\{ [-2(1-y^2)+(1-y)^2]P^{(2,3)}_p+(p+6)(1-y^2)[P^{(2,3)}_p-P^{(2,4)}_p]    \right\}^2\nonumber\\&
=\Bigg\{ [-2(1-y^2)+(1-y)^2]P^{(2,3)}_p+(p+6)(1-y^2)\bigg[  P^{(2,3)}_p-\frac{(2p+6)}{(p+6)}P_p^{(2,3)}\nonumber\\&\hspace{0.5 cm}+\frac{(p+2)}{(p+6)}P_{p-1}^{(2,4)}\bigg]  \Bigg\}^2\nonumber\\&
=\left\{ (1-y)\left[2-(p+3)(1+y)\right]P^{(2,3)}_p+(p+2)(1-y^2)P_{p-1}^{(2,4)}  \right\}^2\nonumber\\&
\label{soa20}
\end{align}
where we have used identities (\ref{soa22}), (\ref{soa23}), (\ref{soa24}) and (\ref{soa25}) in the second, third, fifth and seventh steps of (\ref{soa20}) respectively.

Now, integral $b$ can be rewritten as
\begin{align}
b=a_1\int_{-1}^1(1+y)^3&(1-y)P_{2p+3}^{(2,3)}B(y) dy&\nonumber\\&=a_1\int_{-1}^1(1-y)^{-1}B(y)\frac{d^{2p+3}}{dy^{2p+3}}[(1-y)^{2p+5  }(1+y)^{ 2p+6 }]dy
\label{soa9}
\end{align}
where we use $B(y)$ as given by (\ref{soa20}) and also make the substitution for $P^{(2,3)}_{2p+3}$ using (\ref{soa30}).
Upon doing integration by parts, any term in $(1-y)^{-1}B(y)$ which is a polynomial of degree less than $2p+3$ will have a zero contribution to the integral. In the end, we are left with
\begin{align}
b&=a_1\int_{-1}^{1}(1+y)^3(1-y)P_{2p+3}^{(2,3)}B(y) dy\nonumber\\&=a_1\int_{-1}^1(1+y)^3(1-y)P_{2p+3}^{(2,3)}\left[(p+3)^2(1-y)^2y^2(P_p^{(2,3)})^2\right]dy\nonumber\\&
=a_1\int_{-1}^1(1+y)^3(1-y)^3P_{2p+3}^{(2,3)}(p+3)^2 y^2(P_p^{(2,3)})^2dy\nonumber\\&
\hspace{0.5 cm}-a_1\int_{-1}^1(1+y)^3(1-y)^3P_{2p+3}^{(2,3)}(p+3)^2(P_p^{(2,3)})^2 dy\nonumber\\&
=a_1\int_{-1}^1(1+y)^3(1-y)P_{2p+3}^{(2,3)}\left[(p+3)^2(1-y)^2(y^2-1)(P_p^{(2,3)})^2\right]dy\nonumber\\&
=-a_1\int_{-1}^1(1+y)^4(1-y)^4P_{2p+3}^{(2,3)}(p+3)^2(P_p^{(2,3)})^2dy\nonumber\\&
=-a
\label{soa10}
\end{align}
where we have added a extra term in the third step such since its addition doesn't affect the original integral. This is because using (\ref{soa30}), one can rewrite this term in the following manner:
\begin{align*}
-a_1&\int_{-1}^1(1+y)^3(1-y)^3P_{2p+3}^{(2,3)}(p+3)^2(P_p^{(2,3)})^2 dy\nonumber\\&
=-a_1\int_{-1}^1(p+3)^2(1-y)(P_p^{(2,3)}(y))^2\frac{d^{2p+3}}{dy^{2p+3}}\left[(1-y)^{2p+5}(1+y)^{2p+6}\right]
\label{soa40}
\end{align*}
Similar to (\ref{eq126.11}), performing integration by parts gives rise to boundary terms, which vanish. So, what one is essentially left with is a integral in which the derivative operator acts $2p+3 $ times on a polynomial of degree $2p+1$, which makes it zero.

Thus, the R.H.S. of (\ref{soa3}) is just a constant and the resonant modes are indeed absent for this particular harmonic.

Of course, we have only considered a very specialized case here. We have restricted to the tensor sector and considered possible excitation of a mode with multi-index $\tilde{\textbf{k}}$. There could be excitations of modes with other multi-indices within the tensor sector. Further, tensor harmonics at the linear level would excite scalar and vector harmonics as well. As we go to higher dimensions, the number of possible excitations also increase significantly, as the components of the multi-index increase with dimensions. In general, it would be an arduous task to check for all possible harmonic excitations, even for a single mode data as we go to higher and higher dimensions.

However, it is interesting that in the case considered, namely the tensor sector with multi-index $\tilde{\textbf{k}}$, the resonant term exactly vanishes. Employment of brute force techniques would become increasingly impractical in higher dimensions. It would be exciting if a rigorous proof of absence of resonances at second order were found.
\section{Summary and Discussion}
The main objective of this work was to generalize the study of nonlinear perturbations of anti-de Sitter spacetime to dimensions greater than four. Since the background AdS geometry is spherically symmetric, we use the Kodama-Ishibashi formalism to simplify the working equations and extend its usage to higher orders in perturbation theory. The various metric perturbations are decomposed based on their behaviour on the $n-$ sphere. Since we are dealing with spacetime dimension greater than four, apart from scalar and vector perturbations, we now have tensor perturbations as well. Tensor perturbations, which are the simplest to deal with, are governed by just one equation. Moreover, they are already in aAdS form.

Vector and scalar sector are more involved and one needs to tackle a set of coupled equations at each order to get a single second order equation governing a master variable. Once we solve for this equation,  it is possible to obtain each of the gauge invariant variables solely in terms of the master variable. The next step is to construct metric perturbation, which depend on these gauge invariant variables. But these perturbations do not, by default satisfy aAdS conditions. In order to render them asymptotically AdS, we make use of the gauge freedom. One thing to note is that beyond the linear level, the gauge invariant variables also depend on the source terms, which in turn are composed from metric perturbations of previous orders in perturbation theory. So it is important to take into account their leading order behaviour as well, before doing the gauge fixing. It turns out that as $r\rightarrow\infty$, the source terms in general, fall off like $r^{-(2n+k)}$, where $k$ is some integer. For $n>2$ these sources fall off fast enough so as not to spoil the aAdS structure of the metric perturbations.

We also derive expressions for the $l_s=0,1$ as well as $l_v=1$ mode. These modes which are just gauge at linear level, are physical perturbations at subsequent levels.

So far, analysis of the resonant structure of perturbed equations in four dimensions \cite{diasantos} as well as the five dimensional biaxial Bianchi IX case \cite{adsbi5} had revealed the presence of irremovable resonances only at the third order - resonances are absent at second order.
In this paper, we also study the perturbed equations for special classes of perturbations of the vacuum Einstein equations in dimension five. We examine the resonant structure of equations at the second order, by starting out with only tensor-type perturbations at the linear level. We take single mode initial data corresponding to
multi-index $\tilde{\textbf{k}}$ (given in section 9) and study excitation at second order for a tensor mode with the same multi-index. The resonant terms exactly vanish. While it would be an arduous task to check every tensor mode as well as the scalar and vector sectors, this result is very suggestive. It leads to the interesting problem of whether one can rigorously prove the absence of resonances at second order. We hope to take this up in future work.

\section{Acknowledgement} D.S. Menon thanks the Council of Scientific and Industrial Research (CSIR), India for financial assistance. We thank the referees for their very useful comments.\footnote {In particular, we detected that a numerical factor had been missed in the Mathematica file of an earlier version of this paper. The referees pointed out that resonances were expected to vanish at second order, which led us to examine the file again.}

\section{Appendix}

Here, we list the details pertaining to the angular integral given by (\ref{soa2})
We use the following convention for the $3-$sphere metric:
\begin{align}
d\Omega^2_3=\gamma_{ij}dw^idw^j=d\chi^2+\sin^2\chi(d\theta^2+\sin^2\theta d\phi^2)
\end{align}
$Y^{lm}(\theta,\phi)$ refers to the scalar spherical harmonics.
We use the formulae given in \cite{lindblom} to evaluate the two classes of various traceless, divergence-free tensors.
For our case, since $l=l^{(1)}$, the following expressions hold true
\begin{align}
\mathbb{T}^{ll0}_{(1)ij}&=\sqrt{\frac{(l-1)}{2l}}\Bigg\{ \frac{1}{2}E^{ll}(\bar{D}_iF_j^{l0}+\bar{D}_jF_i^{l0})+\csc^2\chi\left[\frac{1}{2}(l-1)\cos\chi E^{ll}+C^{ll}\right]\nonumber\\&
(F_i^{l0}\bar{D}_j\cos\chi+F_j^{l0}\bar{D}_i\cos\chi)  \Bigg\}
\end{align}
\begin{align}
\mathbb{T}^{ll0}_{(2)ij}=\frac{1}{2(l+1)}\left(\epsilon_i^{\:k\ell}\bar{D}_k\mathbb{T}^{ll0}_{(1)\ell j}+\epsilon_j^{\:k\ell}\bar{D}_k\mathbb{T}_{(1)\ell i}^{ll0} \right)
\end{align}
where for $l=l^{(1)}$,
\begin{align}
C^{ll}=(-1)^{l+1}2^l l!\sqrt{\frac{2(l+1)}{\pi(2l+1)!}}
\end{align}
\begin{align}
E^{ll}=-\frac{2C^{ll}}{l-1}\cos\chi
\end{align}
\begin{align}
F_i^{lm}=\frac{1}{\sqrt{l(l+1)}}\epsilon_i^{\:jk}\bar{D}_j(\sin^l\chi\: Y^{lm})\bar{D}_k\cos\chi
\end{align}
Here, we have used the convention $\epsilon_{\chi\theta\phi}=-\sin^2\chi\sin\theta$.
Thus, the various traceless, divergence-free tensors are
\begin{align}
\mathbb{T}_{\phi\chi}=\frac{\sqrt{6}}{2\pi}\sin^2\chi\sin^2\theta\cos\theta
\end{align}
\begin{align}
\mathbb{T}_{\theta\phi}=-\frac{\sqrt{6}}{2\pi}\sin^3\chi\cos\chi\sin^3\theta
\end{align}
\begin{align}
\mathbb{T}_{\chi\theta}=\sqrt{\frac{2}{3}}\frac{1}{\pi}\sin\chi\cos\chi\sin\theta\cos\theta
\end{align}
\begin{align}
\mathbb{T}_{\theta\theta}=-\frac{1}{\sqrt{6}\pi}\sin^2\chi\left[3\cos^2\chi\sin^2\theta-1\right]
\end{align}
\begin{align}
\mathbb{T}_{\chi\chi}=-\frac{1}{\sqrt{6}\pi}[3\cos^2\theta-1]
\end{align}
\begin{align}
\mathbb{T}_{\phi\phi}=\frac{1}{\sqrt{6}\pi}\sin^2\chi\sin^2\theta\left[\sin^2\theta(1-3\sin^2\chi)+\cos^2\theta\right]
\end{align}

\end{document}